\ifdefined\pdfobjcompresslevel
\fi
\PassOptionsToPackage{table}{xcolor}
\PassOptionsToPackage{twoside=false}{geometry}
\documentclass[manuscript,screen,nonacm]{acmart} 

\AtBeginDocument{%
  }

\usepackage{algorithm}
\usepackage{algpseudocode}
\usepackage{array}
\usepackage{enumitem}
\usepackage{ragged2e}
\usepackage{tabularx}

\definecolor{appendixheader}{HTML}{E6E1D7}
\definecolor{appendixsubheader}{HTML}{F5F2EC}
\definecolor{appendixrule}{HTML}{C8C0B2}
\definecolor{appendixpanel}{HTML}{FBF9F5}

\newcolumntype{L}[1]{>{\RaggedRight\arraybackslash}p{#1}}
\newcolumntype{C}[1]{>{\Centering\arraybackslash}p{#1}}
\newcolumntype{Y}{>{\RaggedRight\arraybackslash}X}

\renewcommand\footnotetextcopyrightpermission[1]{}
\begin{document}

\title[Digital Harf]{Digital Harf: A Clinically Integrated Multimodal AI System for Pervasive Arabic Speech and Language Therapy}

\author{Asif Azad}
\email{asifazad0178@gmail.com}
\authornote{Corresponding author.}
\authornote{These authors contributed equally to this work.}
\affiliation{%
  \institution{Ministry of Defense}
  \city{Riyadh}
  \country{Saudi Arabia}
}

\author{Mohammad Sadat Hossain}
\email{sadat@cse.buet.ac.bd}
\authornotemark[2]
\affiliation{%
  \institution{Ministry of Defense}
  \city{Riyadh}
  \country{Saudi Arabia}
}

\author{MD Sadik Hossain Shanto}
\email{shantosadikrglhs@gmail.com}
\authornotemark[2]
\affiliation{%
  \institution{Ministry of Defense}
  \city{Riyadh}
  \country{Saudi Arabia}
}

\author{Sabri Boughorbel}
\email{sabri.boughorbel@gmail.com}
\affiliation{%
  \institution{Ministry of Defense}
  \city{Riyadh}
  \country{Saudi Arabia}
}

\author{Abdulrhman Aljouie}
\email{aljouie@gmail.com}
\affiliation{%
  \institution{Ministry of Defense}
  \city{Riyadh}
  \country{Saudi Arabia}
}

\author{Bdour Alwuqaysi}
\email{bdour.alwuqaysi@hotmail.com}
\affiliation{%
  \institution{Ministry of Defense}
  \city{Riyadh}
  \country{Saudi Arabia}
}

\author{Yahya Bokhari}
\email{Yahya.bokhari@gmail.com}
\affiliation{%
  \institution{Ministry of Defense}
  \city{Riyadh}
  \country{Saudi Arabia}
}

\author{Ayah Othman Sindi}
\email{asindi@abilitycenter.sa}
\affiliation{%
  \institution{Ability Center}
  \country{Saudi Arabia}
}

\author{Ehsan Hoque}
\email{mehoque@cs.rochester.edu}
\affiliation{%
  \institution{Ministry of Defense}
  \city{Riyadh}
  \country{Saudi Arabia}
}
\affiliation{%
  \institution{University of Rochester}
  \city{Rochester}
  \state{New York}
  \country{USA}
}


\renewcommand{\shortauthors}{Azad et al.}
\begin{abstract}
  Children with Autism Spectrum Disorder in Arabic-speaking countries face
  compounded barriers to effective speech and language therapy: a shortage of
  qualified specialists, limited service reach beyond urban centers, and a
  near-total absence of culturally grounded digital therapy materials. We
  present Digital Harf, a pervasive, multimodal AI platform that extends
  clinician-led speech and language therapy into the home. The system integrates
  three therapeutic modules---language therapy, speech intelligibility, and
  picture description---within a unified workflow that adapts to each child's
  performance over time. To address the critical shortage of Arabic therapy
  content, we introduce an Agentic Synthetic Data Engine (ASDE) that
  automatically generates culturally relevant images, prompts, and language
  tasks guided by explicit therapeutic and cultural criteria. Expert evaluation
  with 13 licensed Speech-Language Pathologists yielded a 90.1\% clinical
  acceptance rate for ASDE-generated content without any manual curation or selection, and strong ratings for cultural
  and linguistic alignment across the full platform. Digital Harf demonstrates
  that AI-driven therapeutic systems can be built from the ground up for
  underrepresented linguistic settings, treating cultural grounding as core
  infrastructure rather than an adaptation afterthought.
\end{abstract}




\maketitle

\section{Introduction}

Autism Spectrum Disorder (ASD) is a neurodevelopmental condition characterized
by challenges in social interaction, communication, and behavior, often
requiring long-term, structured intervention. It is also a growing public
health concern, particularly in regions where infrastructure remain underdeveloped. In Arabic-speaking countries, understanding of ASD prevalence
and service capacity has lagged behind high-resource settings. Reported
prevalence figures across the Arab world vary widely---from below 0.1\% to
above 2\%---largely because most estimates come from clinic- or school-based
studies rather than standardized national
surveillance~\cite{khamees2025estimating}. Even in the Gulf Cooperation Council
(GCC) countries, where screening infrastructure is more developed, officially
reported prevalence has been shown to significantly underestimate the true
burden; recent estimates for one GCC country, Saudi Arabia, place it between 1.7\% and 1.8\%,
comparable to figures from high-surveillance settings such as the United States
(approximately 3.2\%) and the United Kingdom (approximately
1.8\%)~\cite{khamees2025estimating}. Given the population of the Middle East
region, these figures translate to hundreds of thousands of children requiring
varying levels of clinical and developmental support.

Early and continuous intervention can significantly improve developmental
outcomes, particularly in communication and social functioning. Communication and language challenges are central features of ASD for many children, which directly impact a
child's ability to interact, learn, and participate in daily life. This makes
access to effective speech and language therapy essential for long-term quality
of life.

Despite this demand, access to consistent, high-quality therapy remains
limited. The availability of qualified Speech-Language Pathologists (SLPs) in
many Arabic-speaking countries is insufficient to meet population needs, with
services concentrated in major urban centers~\cite{ghali2023audiology}. This
shortage constrains therapy frequency and often leaves caregivers without
structured guidance to support continued practice at home, creating a critical
gap between clinical sessions and everyday life.

This is not only a workforce problem---it is also a content problem that extends well beyond language translation. The vast majority of digital therapy tools
have been designed for English-speaking settings. Their prompts, scenes,
vocabulary, and scenarios do not transfer cleanly to Arabic-speaking children.
A picture or story may look harmless on paper, yet feel unfamiliar because the
objects, places, routines, or wording do not match the child's everyday world.
For clinicians, preparing localized materials by hand is time-consuming,
especially when content needs to be refreshed as the child progresses. This
content gap is a structural barrier that limits the reach and effectiveness of
digital therapy in non-English
settings~\cite{almurashi2022asdreview,lewis2025cld}.

These pressures---workforce scarcity, content misalignment, and the absence of
culturally grounded digital tools---create a strong case for rethinking therapy
delivery through artificial intelligence (AI) and ubiquitous computing. The
goal is not to replace clinician-led intervention, but to extend its reach
into the home while keeping therapy materials fresh, personalized, and
culturally grounded.

In this work, we present \textit{Digital Harf}, a pervasive, multimodal AI
ecosystem designed to complement traditional speech and language therapy for
Arabic-speaking children with ASD. The system was developed through iterative
collaboration with SLPs from multiple clinical centers across the Middle East,
with continuous clinician feedback shaping its design, workflows, and content
generation. The platform integrates multiple therapeutic
modalities---picture description, speech intelligibility, and language
therapy---into a unified, personalized workflow accessible via a web-based
interface.

A key component is the Agentic Synthetic Data Engine (ASDE), a
clinician-informed pipeline that generates culturally relevant images, prompts,
and language tasks guided by explicit therapeutic and cultural criteria. By
automating content creation, ASDE directly addresses the content gap and
enables scalable production of therapy materials without requiring clinicians
to hand-craft every item.

To assess the system, we conduct a preliminary expert evaluation with
practicing SLPs, examining clinical relevance, usability, cultural alignment,
and applicability. While a full clinical trial measuring child-level outcomes
is now underway, this expert-driven assessment provides early evidence of the
system's feasibility for real-world deployment.

The primary contributions of this work are:
\begin{itemize}
    \item \textbf{Pervasive Multimodal AI Therapeutic System.} A web-based ecosystem integrating speech and language therapy into a unified, personalized workflow for continuous home-based intervention, developed through iterative co-design with practicing SLPs.

    \item \textbf{Agentic Synthetic Data Engine (ASDE).} An automated, clinician-informed pipeline for generating culturally relevant therapeutic materials, directly addressing the content gap in Arabic and Middle Eastern contexts.

    \item \textbf{Generalizable Low-Resource Framework.} A replication blueprint with practical guidelines for adapting AI-driven therapeutic systems to other low-resource linguistic settings.
\end{itemize}

\section{Related Works}

We organize related work around three strands: (1) digital tools for speech and
language intervention, with attention to non-English and low-resource settings,
(2) AI-driven interactive systems for autism therapy that involve families
alongside children, and (3) HCI research that treats therapy tools as part of a
broader social arrangement involving clinicians, caregivers, and cultural
context. Reading across all three strands, a structural gap becomes visible:
existing systems address one slice of the therapeutic workflow at a time, are
almost exclusively designed for English-speaking settings, and rarely treat
culturally grounded content generation as a first-class engineering problem.
\textit{Digital Harf} is designed around precisely that gap.

\subsection{Digital Speech and Language Intervention}

Research on digital speech and language intervention has usually moved one piece
at a time. Some systems focus on articulation drills, some on home exercises,
some on telepractice, and others on narrow forms of automated scoring. Even so,
that body of work established something important: repeated therapy becomes much
easier to sustain when it is woven into play, routine, and everyday media use
rather than left as a static worksheet or isolated clinical task. Zajc et~al.
show how tablet-supported speech therapy can be embedded in children's familiar
media practices, while Hair et~al. demonstrate through a longitudinal home
deployment of Apraxia World that a therapy game can remain engaging over time and
meaningfully complement conventional home
practice~\cite{zajc2018tablet,hair2021apraxia}.

Recent reviews make clear, however, that the landscape is still fragmented.
Attwell et~al. survey online speech-therapy systems across home and clinical
contexts and argue that most tools are designed to support clinicians with
specific tasks rather than to offer a broader, connected therapeutic
environment~\cite{attwell2022ostreview}. Deka et~al. review AI-based tools for
speech sound disorder, and Moulaei et~al. map the wider tele-speech literature,
including its interventions, benefits, and delivery
challenges~\cite{deka2025aistreview,moulaei2025telespeech}. More recently, Deka
et~al. advocate for human-centered AI in speech therapy for low-resource
settings, arguing that automated diagnostic tools must account for linguistic
diversity and clinician workflows rather than treating speech as a uniform
signal~\cite{deka2025hcai}. Baird et~al. examine how automatic speech
recognition and assessment models can be incorporated into digital therapeutics
for children with ASD, reporting promising but still limited accuracy when
applied to atypical speech~\cite{baird2024asrdtx}. Taken together, these reviews
show real momentum in digital speech support, but also a field that is still
organized around isolated functions more than around continuous care.

A persistent challenge across this literature is that most speech assessment
tools are built for English. Attia et~al. show that even general-purpose ASR
systems like Whisper suffer a substantial performance gap on children's speech
compared to adults, and that bridging that gap requires child-specific
fine-tuning and data preprocessing~\cite{attia2024kidwhisper}. When the target
language is not English, the problem compounds: Kim et~al. validate a wav2vec
2.0--based ASR system for assessing Korean children with speech sound
disorders, achieving high agreement with clinician transcription, but note that
the approach depends on language-specific annotated corpora that do not yet
exist for most languages~\cite{kim2025koreassr}. For Arabic, the situation is
even more acute. Standardized phonological assessment tools are scarce, and the
few digital systems that exist---such as KidzCare for Arabic speech
screening---remain commercially oriented and clinically
unvalidated~\cite{kidzcare2025app}. Critically, none of these systems address
the upstream content problem: even if an ASR model could score Arabic children's
speech, there is no scalable mechanism for generating the culturally grounded
stimuli that the scoring would require. \textit{Digital Harf}'s Agentic Synthetic Data
Engine (ASDE) is designed precisely to fill that gap, producing clinically
aligned, culturally grounded therapy stimuli and
embedding them within a unified therapeutic workflow.

\subsection{AI-Driven Interactive Systems for Autism Therapy}

A growing body of work has begun to combine AI, augmented reality, and
generative models to create more engaging, personalized therapeutic experiences
for children with ASD. Lyu et~al. developed Eggly, a mobile AR neurofeedback
training game that monitors mu rhythm suppression via a consumer-grade EEG
headband and dynamically adjusts game difficulty to maintain a child's social
attention; evaluated with five children over a ten-month iterative design
process, Eggly demonstrates that combining AR immersion with real-time
brain-state feedback can sustain engagement in therapeutic
play~\cite{lyu2023eggly}. The same group later built EMooly, a tablet-based game
that integrates generative AI for personalized social story creation with
AR-based emotion recognition activities, explicitly involving caregivers in a
collaborative, multi-phase learning process; a controlled study with 24 children
showed significant improvements in emotion recognition compared to a traditional
baseline~\cite{lyu2024emooly}.

At the intersection of AI and clinical dialogue, Lai et~al. introduced
ASD-iLLM, a large language model fine-tuned on clinical intervention transcripts
to conduct ABA-style dialogue with autistic children, demonstrating that LLMs
can be trained to adopt clinically appropriate intervention
strategies~\cite{lai2025asdillm}. Choi et~al. explored how autistic adults
integrate LLM-driven conversational agents into everyday life, finding that
participants valued the non-judgmental interaction space while raising concerns
about the critical evaluation of AI-generated
responses~\cite{choi2024unlocklife}. In a 12-month observational study, Atturu
et~al. evaluated CognitiveBotics, an AI-based therapy platform used alongside
standard care for children with ASD, and reported significant improvements in
receptive and expressive language scores, social maturity, and developmental
quotients for the intervention group~\cite{atturu2025cognitivebotics}.

Perry et~al. offer a broader view in their systematic review of AI-assistive
technologies for adaptive functioning in neurodevelopmental conditions, finding
that while existing tools show promise across social skills, daily living, and
communication domains, the evidence base remains thin and skewed toward
controlled rather than real-world
deployments~\cite{perry2024aireview}. Almurashi et~al. review augmented reality,
serious games, and PECS-based interventions for people with ASD, showing how much
effort has gone into building attractive, child-friendly supports while also
noting how uneven the evidence base remains across modalities and cultural
settings~\cite{almurashi2022asdreview}. A shared limitation runs through this
body of work: each system optimizes a single interactive modality---emotion
recognition, social story generation, ABA dialogue, or neurofeedback---and none
addresses the upstream question of how clinically appropriate, culturally
situated content should be produced at scale for non-English populations.
\textit{Digital Harf} targets this gap directly, integrating content generation,
multimodal therapy delivery, and caregiver-facing interpretation within a single
Arabic-first workflow.

\subsection{Therapy as a Sociotechnical Arrangement}

HCI and autism research adds another important perspective by treating therapy
tools as part of a larger social arrangement rather than as standalone
automation. Porayska-Pomsta et~al. showed that blending human and artificial
intelligence could support autistic children's social communication in real
educational settings, emphasizing that use practices matter as much as the
technical system itself~\cite{porayskapomsta2018blending}. Koronth\'{a}ly
et~al. trace the co-evolution of language, technology, and culture across two
decades of autism computing research in the ACM Digital Library, showing how the
field's vocabulary and design assumptions have shifted and arguing that future
systems must be more reflexive about the cultural frames they
embed~\cite{koronthaly2025coevolution}. Marcu et~al. conducted an early field
study in which parents of children with ASD used wearable cameras at home,
finding that parent-driven capture surfaced therapeutically relevant moments for
clinician review, but also highlighted privacy and adoption challenges in
home-deployed technology~\cite{marcu2012wearablecameras}.

More recent work has become even more family centered. AACessTalk supports
communication between minimally verbal autistic children and parents through
contextual guidance and card recommendation~\cite{choi2025aacess}. Dangol
et~al. examine AI-supported home practice as a way of helping caregivers think
more like speech-language pathologists during everyday
practice~\cite{dangol2025iwantslp}. Chueh et~al. report a pilot generative-AI
language-therapy platform for children with language delay that combined home
practice, caregiver monitoring, and interactive exercises, with stronger gains
among heavier users~\cite{chueh2025genai}. In parallel, Suh et~al. examine
opportunities and challenges in AI support for speech-language pathologists,
while Lewis et~al. show why tools for culturally and linguistically diverse
children cannot be treated as culturally neutral by
default~\cite{suh2024opportunities,lewis2025cld}. These contributions are
important, but they share a common ceiling: each covers one segment of the
therapy arc---home engagement, caregiver scaffolding, or clinician
support---without addressing how culturally appropriate content enters the
system in the first place, or how the full chain from content generation to
clinician interpretation to caregiver follow-through can be held together in a
single, non-English-first platform.

\subsection{Positioning Digital Harf}

The three strands reviewed above converge on a revealing pattern: existing
systems are sophisticated within their chosen scope, but that scope is almost
always narrow, English-centric, and upstream-agnostic. Digital speech tools
automate scoring or support home practice, but do not generate the stimuli that
scoring and practice require. AI-driven autism therapy systems improve
engagement and personalization, but presuppose a ready supply of culturally
appropriate content. Sociotechnically informed tools bring caregivers and
clinicians closer together, but stop short of addressing the content and
language gaps that make those tools inaccessible to Arabic-speaking families in
the first place. Three specific deficiencies stand out across all three strands.
First, no existing system treats culturally grounded content generation as a
core engineering problem: materials are assumed to exist and are not produced by
the platform itself. Second, no system integrates content creation,
multimodal therapy delivery, and caregiver-facing clinical interpretation
within a single, coherent workflow. Third, no system has been designed from the
ground up for Arabic, a language with distinct phonological and morphological
properties and a large, underserved population of children with ASD who are
currently without adequate digital support.

\textit{Digital Harf} is designed to address all three deficiencies simultaneously.
Its Agentic Synthetic Data Engine (ASDE) generates clinically aligned, culturally
grounded stimuli for Arabic phonemes at scale, treating content production as an
automated, clinician-informed pipeline rather than a manual preparation task.
Its unified web-based platform connects that content to multiple therapeutic
modalities---picture description, speech intelligibility, and language
therapy---within a single personalized workflow. And its caregiver-facing
interface closes the loop between clinical sessions and home practice, giving
non-specialist caregivers structured, interpretable guidance without requiring
an SLP to be present. Developed through iterative co-design with practicing
SLPs across multiple clinical centers in the Middle East, and evaluated through
expert clinical assessment prior to a full child-level trial, \textit{Digital Harf}
represents not just another therapy application but a connected multimodal
infrastructure built for a setting where specialist time, validated assessment
tools, and culturally matched digital resources are all simultaneously scarce.
In doing so, it also offers a replicable blueprint for other low-resource
linguistic communities facing the same structural barriers.
\section{Design Context and Requirements}\label{sec:design-context}

This section describes the conditions that shaped \textit{Digital Harf}'s design and
the four requirements that emerged from them.

\textit{Digital Harf} grew out of a simple problem: speech and language therapy works
best when children can practice often, but many Arabic-speaking families cannot
access that level of support through clinic visits alone. Across the Middle
East, the number of children who may need autism-related support is likely much
higher than older official figures suggested~\cite{khamees2025estimating}, yet
access to therapy remains shaped by limited specialist capacity, uneven service
distribution, and the difficulty of sustaining frequent in-person care.

Compounding this access problem is a content problem that affects every
Arabic-speaking context. Most digital therapy tools---from interactive apps to
assessment instruments---were designed for English-speaking children in Western
cultural settings. Their prompts, scenes, vocabulary, and examples do not
transfer cleanly. A language therapy card asking a child to identify ``a
mitten'' or a scene depicting a Thanksgiving dinner is not a translation
problem but a fundamental mismatch in cultural reference. For Arabic-speaking
children, effective content must reflect their lived environment---local foods,
familiar clothing, everyday routines, and social interactions that resonate
with their daily experience~\cite{lewis2025cld}. For therapists, preparing
such localized materials by hand is time-consuming, especially when content
needs to be refreshed as the child
progresses~\cite{almurashi2022asdreview}.

These conditions shaped \textit{Digital Harf} as a home-extendable system that stays
close to clinical practice instead of trying to replace it. Throughout the
design process, we worked iteratively with practicing SLPs across multiple
centers, incorporating their feedback to ensure that the system's therapeutic
logic, content, and interaction flows aligned with real clinical needs.

\subsection{Design Requirements}

From this context, we distilled four requirements that guided the design.

\paragraph{R1. Clinical Fit and Therapist Alignment.}
Tasks had to match familiar speech and language goals, support multiple
communication skills, and produce performance information that remained
clinically meaningful even when the child was practicing outside the clinic.

\paragraph{R2. Cultural and Linguistic Grounding.}
The platform needed to present children with content that felt genuinely
grounded in their cultural context---not merely translated. Images, prompts,
and scenarios had to reflect locally recognizable settings, objects, routines,
and social norms so children could focus on the therapeutic task rather than on
decoding the context.

\paragraph{R3. Child-Friendly Home Use with Caregiver Support.}
The system needed clear interaction flows, visually supportive screens, and
low-friction tasks that children could complete with caregiver support.
Progress views had to help parents understand what happened in a session and
what to focus on next.

\paragraph{R4. Personalization that Stays Useful Over Time.}
The platform needed to adapt to different skill levels and remain useful for
repeated practice. This shaped the need for a content engine that could keep
producing new materials without losing therapeutic clarity or cultural fit;
without such an engine, localized content would run dry after a few sessions.

These requirements pushed us toward a connected therapeutic ecosystem rather
than a single AI feature, and explain why \textit{Digital Harf} depends so strongly on
its synthetic data engine.

\section{Digital Harf}

This section presents the \textit{Digital Harf} platform and its four core components:
Language Therapy, Speech Intelligibility, Picture Description, and an AI
Caregiver Assistant. We close with the personalization framework that adapts
content and feedback across modules.

\subsection{System Overview}

\textit{Digital Harf} is a pervasive, multimodal AI ecosystem designed to complement
traditional speech and language therapy for Arabic-speaking children with
Autism Spectrum Disorder (ASD). The system enables structured, continuous
practice beyond clinical settings by supporting interactions among three key
stakeholders: children, caregivers, and Speech-Language Pathologists (SLPs).
Through a web-based interface accessible across devices, \textit{Digital Harf}
facilitates therapy activities in home environments while maintaining alignment
with clinical objectives.

At its core, the platform integrates three primary therapeutic
modules—\textit{Language Therapy}, \textit{Speech Intelligibility}, and
\textit{Picture Description}—each targeting complementary aspects of speech and
language development. These modules are unified within a single workflow that
allows children to engage in guided tasks, receive immediate feedback, and
progressively build communication skills. Unlike conventional digital tools,
\textit{Digital Harf} is not a static application but a dynamic service that adapts to
each child's developmental trajectory through a personalized, data-driven
workflow, enabling both therapist-directed use and independent home practice
supported by caregivers.

Beyond child interaction, \textit{Digital Harf} incorporates a 24/7 AI-driven caregiver
assistant that provides contextual, evidence-based guidance to both parents and
clinicians, supports daily therapy routines, and helps bridge the gap between
isolated clinical sessions and consistent home-based reinforcement. This
component extends the system's utility by reducing caregiver uncertainty and
enabling more effective participation in the therapeutic process.

Underlying the platform is a scalable backend that manages user interactions,
processes multimodal inputs (e.g., speech and text), and delivers real-time
feedback. The system is further supported by a content generation pipeline that
produces culturally relevant therapeutic materials, ensuring that tasks remain
meaningful within Arabic-speaking contexts. Overall, \textit{Digital Harf} functions as
an integrated ecosystem that bridges the gap between limited in-clinic therapy
and the need for continuous, personalized intervention in everyday
environments.

\subsection{Language Therapy}

The \textit{Language Therapy} module is designed to support the development of
foundational receptive and expressive language skills, which are central to
communication in children with Autism Spectrum Disorder (ASD). Grounded in
established speech and language therapy practices, the module targets core
domains such as vocabulary, semantic understanding, categorization, and
functional language use. By structuring tasks across both understanding
(receptive) and production (expressive), the system enables children to engage
with language at multiple levels of complexity while maintaining alignment with
therapist-defined goals.

Interaction with the module follows a structured yet flexible session flow.
Upon entering the module, the user is first presented with a choice between two
modes: \textit{Receptive} and \textit{Expressive}. Within each mode, a set of
domain categories is displayed (e.g., shapes, common objects, actions,
categories), from which one or more domains can be selected. Each selected
domain contributes a fixed number of randomized questions (five per domain),
allowing sessions to scale dynamically based on selection. For example,
selecting two expressive domains and one receptive domain results in a
15-question session (3 domains $\times$ 5 questions each).

\begin{figure}[t]
    \centering
    \includegraphics[width=\linewidth]{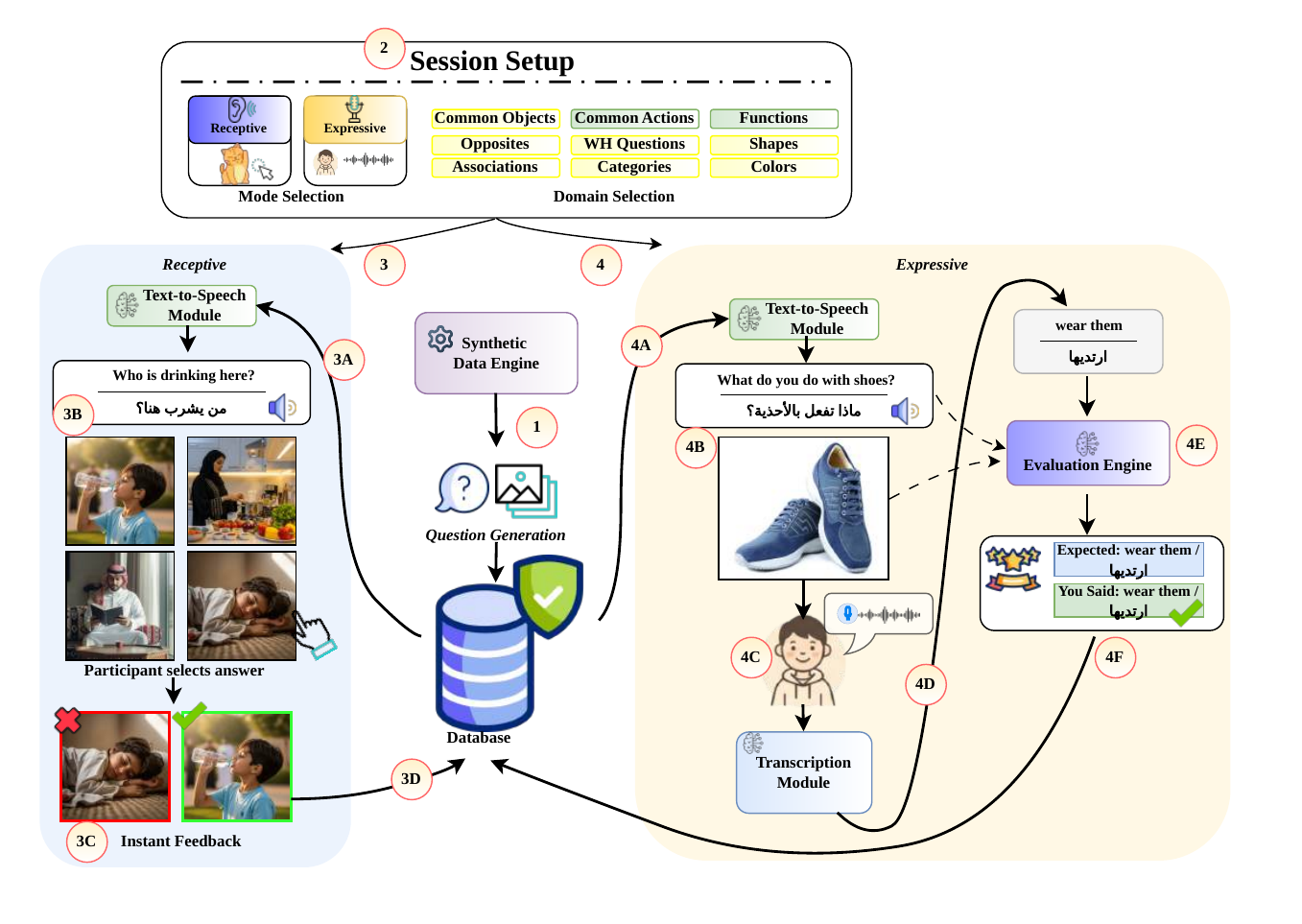}
    \caption{Technical workflow of the Language Therapy module. Therapy content is sourced from an ASDE-populated database. In Receptive mode, user selections are compared against correct answers. In Expressive mode, spoken responses are transcribed and evaluated by a semantic comparison engine that accepts synonyms and paraphrases. Results are returned as immediate feedback and logged for longitudinal tracking. An example full interaction flow is illustrated in Appendix
Figure~\ref{fig:appendix-lt-ui}.}
    \label{fig:lt_pipeline}
    \Description{A flowchart showing the Language Therapy module pipeline. An ASDE content database feeds therapy questions and images into the system. The flow splits into Receptive and Expressive paths, both converging on a feedback and logging step that returns results to the user and stores data for personalization.}
\end{figure}

Once the session begins, the interaction proceeds sequentially. In Receptive
mode, each task presents a question both visually (text) and aurally via a
text-to-speech (TTS) module, ensuring accessibility for children with varying
reading abilities. The child is shown multiple image-based options and selects
the one that best matches the prompt. In Expressive mode, the system similarly
presents a question through both text and audio, after which the child records
a spoken response using a microphone interface. Across both modes, the
interface is intentionally simple and consistent to support independent or
caregiver-assisted use in home environments. Immediate feedback is provided at
the interaction level (e.g., visual indicators for correctness), while a
comprehensive performance summary—including overall accuracy, domain-level
breakdown, and answer review—is presented at the end of the session. This full
interaction flow, including key interface states, is illustrated in Appendix
Figure~\ref{fig:appendix-lt-ui}.

Under the hood, the Language Therapy module operates as a closed-loop system
integrating content generation, multimodal interaction, and automated
evaluation, as shown in Figure~\ref{fig:lt_pipeline}. All therapeutic tasks are
sourced from a central database populated by the Agentic Synthetic Data Engine
(ASDE), which generates culturally grounded prompts and associated visual
materials across predefined language domains.

In Receptive mode, prompts are delivered through both text and TTS, and user
selections are evaluated against predefined ground-truth labels. In Expressive
mode, spoken responses are captured and passed through an automatic speech
transcription component. The resulting text is then processed by a semantic
evaluation engine that compares the user's response with expected answers.
Rather than relying on strict string matching, the system accepts semantically
equivalent variations (e.g., synonyms or paraphrased responses), enabling more
robust and realistic evaluation aligned with natural language use. The
evaluation outcomes are returned to the user and logged in the system,
supporting both immediate feedback and longitudinal tracking.

By combining structured interaction design with scalable content generation and
semantically aware evaluation, the Language Therapy module supports continuous,
home-based language practice while remaining grounded in clinically meaningful
assessment criteria.

\subsection{Speech Intelligibility}

The \textit{Speech Intelligibility} module focuses on improving the clarity,
accuracy, and fluency of speech production. It is designed for children who can
produce speech but exhibit articulation errors, phoneme-level difficulties, or
reduced intelligibility. The module provides structured, repeatable speaking
tasks with immediate, fine-grained feedback, enabling progressive refinement of
pronunciation and increased confidence in verbal communication.

\begin{figure}[t]
    \centering
    \includegraphics[width=1.0\linewidth]{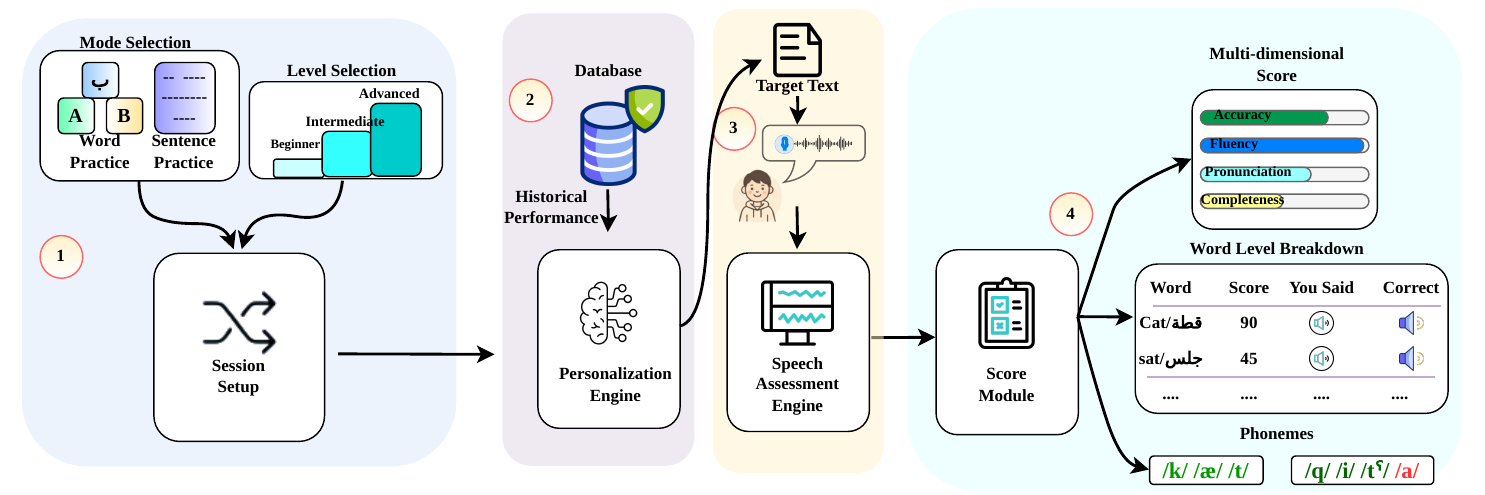}
    \caption{Technical workflow of the Speech Intelligibility module. A Personalization Engine selects five target items per session based on difficulty level and historical performance, prioritizing previously mispronounced items. The user's spoken attempt is processed by a Speech Assessment Engine and scored across accuracy, fluency, pronunciation quality, and completeness. Sentence-level tasks include word-by-word breakdowns. If the score is below threshold, up to two retries are allowed. All data is logged for adaptive future sessions. Appendix Figure~\ref{fig:appendix-si-ui} illustrates an example user-facing interaction flow.}
    \label{fig:si-technical}
    \Description{A flowchart showing the Speech Intelligibility pipeline. A Personalization Engine selects items, the user records speech, a Speech Assessment Engine evaluates it, and a Scoring Module computes multi-dimensional metrics. A threshold check governs retries. All data is logged back for session adaptation.}
\end{figure}

The interaction flow is organized as a structured session to ensure consistency
and measurable progress. The user selects the \textit{Speech Intelligibility}
module and chooses (i) a practice mode—\textit{Word Level} or \textit{Sentence
    Level}—and (ii) a difficulty level (e.g., Easy, Medium, Hard), which controls
the complexity of vocabulary and phoneme combinations.

Based on these inputs, the system initiates a session consisting of five items.
For each item, the user is presented with a target word or sentence along with
a supporting visual cue. An optional audio model can be played via a
text-to-speech interface, enabling multimodal exposure to correct
pronunciation. The user then records their response through the microphone
interface and may replay their own audio for self-assessment.

Immediately after each attempt, the system provides real-time feedback in the
form of an intelligibility score and visual indicators. A threshold-based
mechanism governs progression: if the score exceeds a predefined threshold
(e.g., 80\%), the system advances to the next item; otherwise, up to two
additional attempts are provided. In sentence-level tasks, feedback includes
both an overall score and a detailed word-level breakdown. After completing all
five items, the system presents a session summary highlighting overall
performance, number of attempts, and time spent, enabling caregivers and
clinicians to monitor progress.

Figure~\ref{fig:si-technical} illustrates the system pipeline. The process
begins with \textit{session configuration}, where user-selected parameters
(mode and difficulty) are processed by a \textit{Personalization Engine}. This
component dynamically constructs each session by selecting five target items
based on both the current configuration and the user's historical performance
stored in a central database. This ensures that each session is adaptive,
non-repetitive, and aligned with the user's evolving proficiency.

For each item, the system delivers a \textit{target text} (word or sentence),
accompanied by synthesized audio. The user's spoken response is captured and
processed by a \textit{Speech Assessment Engine}, which performs real-time
comparison between the recorded audio and the reference pronunciation.

The assessment engine is designed as a flexible, modular component that
supports plug-and-play integration of different speech processing backends. In
our implementation, we incorporate both a proprietary cloud-based solution
(Azure Cognitive Services~\cite{microsoft_pronunciation_assessment}) and a custom fine-tuned pipeline
(\textit{Harf-Speech}~\cite{azad2026harf}), enabling robust evaluation while
retaining adaptability to language- and domain-specific requirements. This
plug-and-play design enables flexibility across languages and deployment
settings, particularly in low-resource contexts where domain-specific
adaptation may be required.

The assessment output is passed to a \textit{Scoring Module}, which computes
multi-dimensional metrics including accuracy, fluency, pronunciation quality,
and completeness. In addition to an overall score, the system generates a
detailed word-level breakdown, enabling identification of specific errors.
Where supported, phoneme-level analysis further decomposes speech into
constituent sound units, allowing fine-grained articulation feedback.

All interaction data, including scores, attempts, and error patterns, are
logged back into the database. This closed-loop design enables continuous
adaptation: subsequent sessions prioritize previously mispronounced words and
challenging phonemes, ensuring targeted and efficient practice. By combining
structured interaction, real-time assessment, and adaptive content selection,
the module supports sustained improvement in speech intelligibility within
home-based environments.

\subsection{Picture Description}

The \textit{Picture Description} module focuses on developing expressive
language, narrative ability, and context-driven communication. It is designed
for children who can produce words or short phrases but struggle to construct
complete, meaningful descriptions. The module emphasizes structured
storytelling through visual stimuli, enabling children to progressively improve
vocabulary usage, grammatical structure, and content richness within realistic
scenarios.

\begin{figure}[t]
    \centering
    \includegraphics[width=0.9\linewidth]{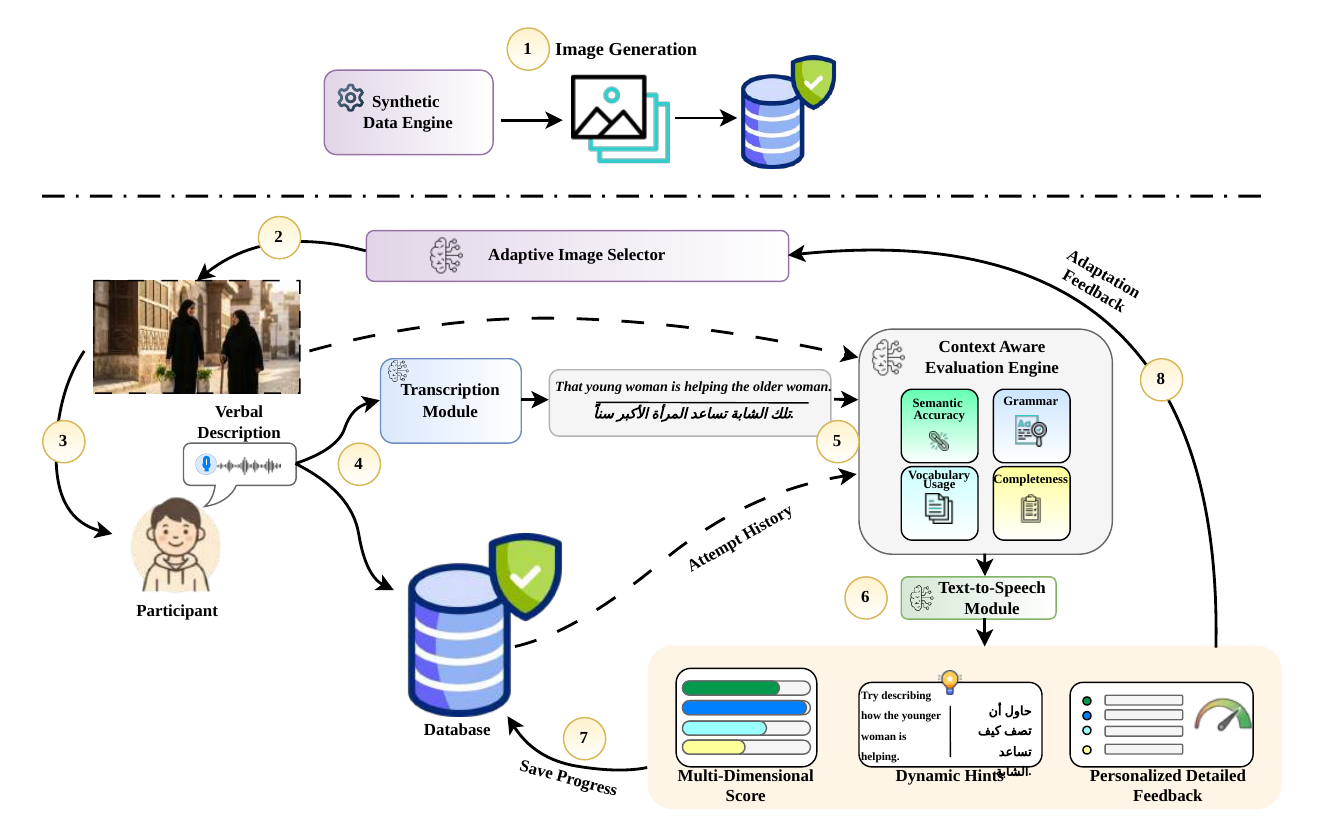}
    \caption{Technical workflow of the Picture Description module. An Adaptive Image Selector retrieves images based on difficulty level and past performance. The user's spoken description is transcribed and analyzed by a Context-Aware Evaluation Engine across four dimensions: semantic accuracy, grammatical correctness, vocabulary richness, and content completeness. Early attempts produce targeted suggestions; the final attempt yields a comprehensive evaluation report. All data is logged for adaptive future sessions. Appendix Figure~\ref{fig:appendix-pd-ui} illustrates an example user-facing interaction flow.}
    \label{fig:pd-technical}
    \Description{A flowchart showing the Picture Description pipeline. A Synthetic Data Engine populates a therapy image database. An Adaptive Image Selector retrieves images. The user records a description, which is transcribed and analyzed by a Context-Aware Evaluation Engine. Suggestions or evaluation reports are generated. All data is logged for future adaptation.}
\end{figure}

The interaction is designed as a guided, multi-attempt session centered around
a single image. The user selects the \textit{Picture Description} module and
chooses a difficulty level (e.g., Basic or Advanced), which controls the
complexity of the visual scene. The system then presents a culturally relevant
image generated and curated through the platform's Agentic Synthetic Data
Engine (ASDE; Section~\ref{sec:sde}).

The user begins by observing the image and recording a verbal description using
the microphone interface. The interaction is structured around a three-attempt
cycle. After the first attempt, the system provides immediate, actionable
suggestions (e.g., prompting the user to include missing details such as
objects, actions, or relationships). The user then records a second attempt,
incorporating the feedback, followed by refined suggestions from the system.

After the third attempt—or earlier if the user chooses to proceed—the system
provides a comprehensive feedback report. This report evaluates the response
across multiple dimensions, including semantic correctness, vocabulary quality,
grammatical structure, and content completeness. Each dimension is presented
with a numerical score and targeted feedback, allowing the user to understand
both strengths and areas for improvement. Throughout the interaction, the user
can replay their recordings, enabling self-assessment and reflection.

Figure~\ref{fig:pd-technical} illustrates the underlying system pipeline, which
operates as a feedback-driven interaction loop rather than a single-pass
process. The workflow begins with the \textit{Synthetic Data Engine}, which
generates culturally grounded images aligned with therapeutic goals and stores
them in a central database. 
At runtime, an \textit{Adaptive Image Selector} retrieves an appropriate image for each session through the \textit{Score-Aware Matching (SAM)} algorithm (see Appendix Section \ref{sec:appendix-sam}, Algorithm \ref{alg:sam} for the full pseudocode). 
Rather than serving images at random, SAM analyzes the
authenticated user's full session history and partitions the available image
pool into four tiers: an \textit{initial} pool of images the user has never
encountered, a \textit{hard} pool of lowest-scoring images, a \textit{medium}
pool, and an \textit{easy} pool of highest-scoring images. Recently seen images
are excluded to prevent repetition, with this constraint relaxed automatically
if pools become empty. SAM then probabilistically samples across these tiers,
assigning the highest weight to unseen images to ensure progressive coverage of
new material, while still cycling back to challenging images to reinforce weak
areas. The user's spoken description is captured and passed to a
\textit{Transcription Module}, which converts speech into text. This
transcription, along with prior attempt history, is processed by a
\textit{Context-Aware Evaluation Engine}.

The evaluation engine performs multi-dimensional analysis across four core
aspects: semantic accuracy (alignment with the image), grammatical correctness,
vocabulary richness, and content completeness. Importantly, the engine
incorporates attempt history, allowing it to generate feedback that is
contextually aware of what has already been described or omitted in previous
attempts.

Based on this analysis, the system generates two types of outputs. For early
attempts, it produces \textit{dynamic hints}—targeted suggestions designed to
guide the next response (e.g., prompting additional details or relationships).
For the final attempt, it produces a \textit{comprehensive evaluation report},
including scores and detailed feedback across all dimensions. These outputs are
delivered through both text and optional text-to-speech, supporting multimodal
interaction.

All interaction data, including transcriptions, scores, and feedback history,
are logged back into the database. This enables a closed-loop adaptive system,
where future image selection and feedback strategies are informed by prior
performance. By combining iterative feedback, multi-dimensional evaluation, and
culturally grounded content, the module supports progressive development of
narrative and expressive language skills in home-based settings.

\subsection{AI Caregiver Assistant}\label{sec:ai-assistant}

The \textit{AI Caregiver Assistant} extends \textit{Digital Harf} beyond direct child
interaction by providing continuous, context-aware support to caregivers and
clinicians. While the core therapeutic modules focus on skill development, this
component addresses a critical gap: helping caregivers interpret progress,
reduce uncertainty, and make informed decisions for home-based practice.

Caregivers access the assistant through a bilingual chat interface (Arabic and
English). Unlike generic chatbots, responses are personalized and grounded in
the child's actual performance data across modules. For example, queries about
progress or areas of improvement are answered using the participant's
historical scores, attempt patterns, and error trends, resulting in targeted,
actionable recommendations rather than generic advice. The assistant also
supports general informational queries while clearly indicating when
professional consultation is required.

At the technical level, each query passes through a safety guardrail before
entering an \textit{Agentic Reasoning Core} that plans how to answer by
retrieving relevant data from module-specific performance streams. This data is
analyzed and synthesized into a grounded response, ensuring outputs are
directly supported by evidence rather than heuristic generation. The full
technical architecture, including the guarded reasoning pipeline and data
retrieval flow, is detailed in Appendix Section~\ref{sec:appendix-ai-assistant}
and Figure~\ref{fig:ai-technical}.

\subsection{Personalization}

Personalization in \textit{Digital Harf} is not implemented as a single feature but as a
system-wide design principle that adapts content, feedback, and interaction
flow based on each child's evolving performance. Unlike static therapy tools,
the platform continuously closes the loop between user interaction and
subsequent task generation, enabling sustained, individualized progression
across modules.

At the interface level, the \textit{Dashboard} acts as a central
personalization hub, aggregating session history, performance trends, and
module-specific outcomes into an interpretable view for caregivers and
clinicians. Rather than presenting raw scores, it emphasizes longitudinal
patterns—such as improvement trajectories, consistency, and weak
areas—supporting informed decision-making about future practice.

Within modules, personalization operates at multiple stages. In \textit{Speech
    Intelligibility}, session generation dynamically adapts based on prior
performance, selecting words or sentences that target previously mispronounced
phonemes while adjusting difficulty levels over time. Feedback is similarly
personalized, providing multi-dimensional scoring (e.g., accuracy, fluency,
pronunciation) and highlighting specific articulation issues to guide targeted
improvement.

In \textit{Picture Description}, personalization is driven by a closed-loop
content selection mechanism. The system not only adapts feedback across
attempts—progressing from lightweight hints to comprehensive evaluation—but
also determines future task difficulty through a performance-aware image
selection process. Central to this is the \textit{Score-Aware Matching (SAM)}
algorithm, which operates across frontend-backend interactions. When a new task
is requested, the backend analyzes the authenticated user's historical
performance across all prior sessions, computes per-image average scores, and
partitions content into multiple pools: unseen (initial), hard (low-scoring),
medium, and easy (high-scoring). Recently viewed items are filtered to avoid
repetition. The next image is then selected probabilistically, prioritizing
unseen content while still revisiting challenging items to reinforce learning.
After each attempt, detailed performance data—including scores, transcriptions,
and feedback dimensions—is logged and fed back into the selection process,
enabling continuous adaptation.

Personalization also extends to feedback generation. In \textit{Picture
    Description}, the system provides iterative, context-aware suggestions across
attempts, while in \textit{Speech Intelligibility}, it surfaces fine-grained
error patterns. This ensures that feedback is not only immediate but also
aligned with the child's specific weaknesses rather than generic correctness
signals.

Finally, the \textit{AI Caregiver Assistant} represents personalization at the
conversational level. Unlike generic chat systems, it grounds its responses in
the child's actual performance data across modules, synthesizing insights into
actionable recommendations. This enables caregivers to receive guidance that is
directly relevant to their child's developmental trajectory.

Together, these components form a multi-layered personalization framework that
spans content selection, feedback generation, progress tracking, and caregiver
support. By tightly integrating data collection with adaptive decision-making,
\textit{Digital Harf} ensures that therapy remains responsive, non-repetitive, and
aligned with each child's individual learning needs over time.

\section{Agentic Synthetic Data Engine (ASDE)}\label{sec:sde}

This section describes the Agentic Synthetic Data Engine, \textit{Digital Harf}'s
pipeline for producing culturally grounded therapy materials at scale. ASDE
addresses the content gap through two parallel pipelines---one for Picture
Description scenes and one for Language Therapy cards---each built around the
same agentic architecture of planning, constrained generation, multi-judge
review, and long-term memory for diversity.

\textit{Digital Harf} depends on a steady supply of therapy materials, but these
materials have to satisfy more than visual quality. Picture description scenes
need to be rich enough to invite narration. Language therapy items need clear
answer structure, age-appropriate wording, and clean visual choices. Across
both pipelines, the content also has to remain culturally familiar, safe for
children, and varied enough for repeated use.

We found early on that one-shot generation was not enough. A model could
produce an image that looked realistic but was weak as a therapy prompt. It
could generate a valid question but pair it with confusing distractors. It
could also repeat the same kinds of scenes, objects, and answer patterns too
often. ASDE was built to address these problems. We use the term \emph{agentic}
to describe a workflow in which planning, generation, memory, and review play
separate roles instead of being collapsed into a single prompt.

Across both pipelines, ASDE follows the same basic idea: define the therapeutic
goal first, generate only within clear constraints, review outputs against
explicit quality checks, and keep memory of what has already worked so later
runs stay varied. In practical terms, the question behind both pipelines is
simple: would a therapist feel comfortable using this item in a real session?
As it can be understood, the system is not purely linear. Generation, judging,
and memory interact through repeated feedback loops.

\subsection{Picture Description Pipeline}

\begin{figure*}[t]
    \centering
    \includegraphics[width=\textwidth]{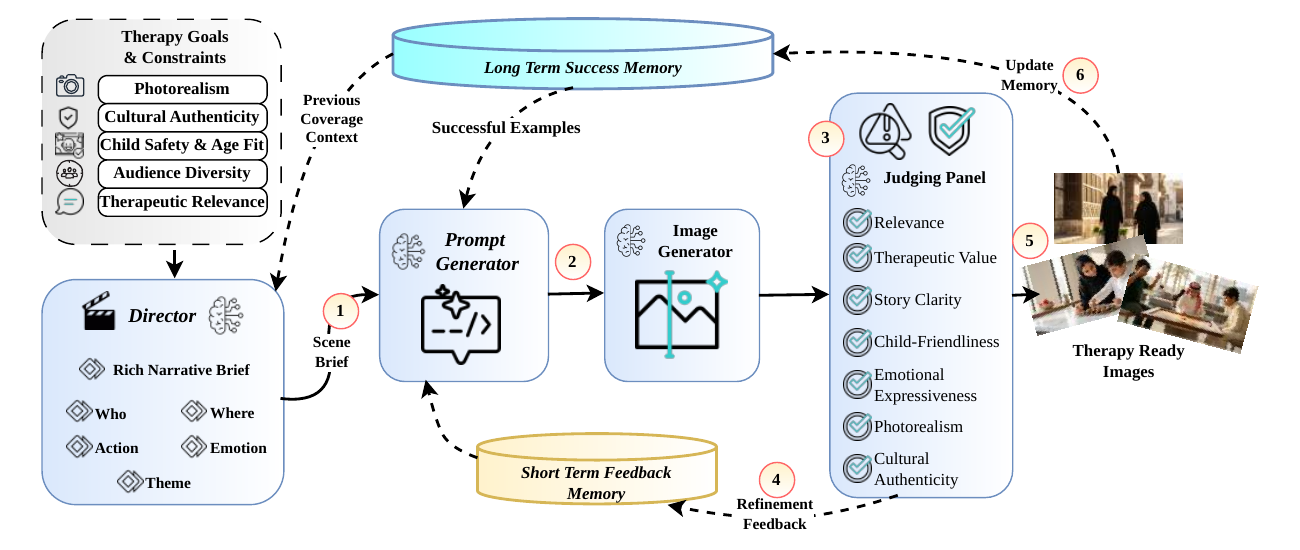}
    \caption{Architecture of the Agentic Synthetic Data Engine for Picture Description (ASDE-PD). The pipeline generates photorealistic images of culturally relevant scenes that children can describe verbally during speech therapy. Therapy goals and cultural constraints (listed on the left---e.g., photorealism, cultural authenticity, child safety) feed into a Director component, which composes a narrative brief specifying the characters, setting, action, emotion, and theme of each scene. A Prompt Generator then expands this brief into a detailed image-generation prompt, drawing on examples of previously successful images stored in long-term memory and---on retry attempts---refinement feedback from short-term memory. The Image Generator renders a candidate image, which a Judging Panel of three independent judges evaluates against seven criteria: relevance to the scene, therapeutic value for description tasks, story clarity, child-friendliness, emotional expressiveness, photorealism, and cultural authenticity. All three judges must independently pass the image with a minimum score of 7.0 out of 10 on each criterion. Accepted images join the therapy-ready library and update long-term memory; rejected images trigger targeted prompt refinement (up to three attempts). The Director reads coverage context from long-term memory to ensure successive batches of images span diverse settings, characters, and themes rather than repeating similar content.}
    \Description{A pipeline diagram for the ASDE picture-description system. On the left, a dashed box lists therapy goals and constraints: photorealism, cultural authenticity, child safety and age fit, audience diversity, and therapeutic relevance. Below it, the Director component produces a rich narrative brief specifying who, where, action, emotion, and theme. A cylindrical database at the top labeled Long Term Success Memory provides previous coverage context to the Director and successful examples to the Prompt Generator. The Director sends a scene brief (step 1) to the Prompt Generator, which outputs to the Image Generator (step 2). The generated image is evaluated by the Judging Panel (step 3), which checks relevance, therapeutic value, story clarity, child-friendliness, emotional expressiveness, photorealism, and cultural authenticity. Rejected images produce refinement feedback (step 4) stored in Short Term Feedback Memory and returned to the Prompt Generator. Accepted images become therapy-ready outputs (step 5) and update long-term memory (step 6).}
    \label{fig:sde-pd}
\end{figure*}

The picture description pipeline produces open scenes that children can
describe in their own words. The task is more demanding than it appears: an
image that looks visually pleasant can still be a weak therapy prompt if the
scene is cluttered, culturally unfamiliar, or too flat to invite narration.
ASDE-PD addresses this through five interacting components shown in
Figure~\ref{fig:sde-pd}: a \emph{Director} that plans scene diversity, a
\emph{Prompt Generator} that translates plans into detailed image briefs, an
\emph{Image Generator} that renders candidate images, a \emph{Judging Panel}
that evaluates them against explicit therapy criteria, and two \emph{Memory
    Layers} that link one generation run to the next.

\subsubsection{Director}

The Director is the creative planning layer of the pipeline. Its role is to
turn high-level therapy goals and cultural constraints---photorealism, cultural
authenticity, child safety and age fit, audience diversity, and therapeutic
relevance---into a batch of rich narrative briefs before any image is rendered.
For each brief, the Director specifies five structural elements: the \emph{who}
(characters), \emph{where} (setting), \emph{action}, \emph{emotion}, and
\emph{theme}. These are not keywords. Each brief is a narrative vignette of
several sentences, providing enough scene structure to anchor a distinctive,
story-rich image rather than a generic scene.

To maintain variety across batches, the Director consults \emph{previous
    coverage context} retrieved from long-term memory before generating each new
set of briefs. This prevents the library from converging on the same character
types, locations, and activities. Within each batch, all images are generated
in parallel, keeping the pipeline efficient at scale.

\subsubsection{Prompt Generation}

The Prompt Generator receives the Director's scene brief and expands it into a
detailed image generation prompt. This two-stage design separates creative
planning from prompt engineering: the Director decides \emph{what} to depict,
while the Prompt Generator decides \emph{how} to describe it to the image
model. The generator enforces all fixed constraints---cultural dress codes,
architectural authenticity, absence of written text, and appropriate age
representation---while remaining faithful to the specific scene the Director
specified.

When an image fails review and the pipeline retries, the Prompt Generator
additionally receives \emph{refinement feedback} from short-term memory. This
allows the prompt to change in a targeted way on the next attempt rather than
regenerating from scratch. On every attempt, the generator is also grounded by
a small set of \emph{successful examples} retrieved from long-term memory,
which help calibrate the level of detail and visual specificity that has worked
in previous runs.

\subsubsection{Image Synthesis}

The expanded prompt is passed to the image generator, which renders a
photorealistic candidate image. Images within each Director batch are generated
in parallel to reduce overall pipeline runtime. Cultural and safety
requirements are enforced at the prompt level rather than through post-hoc
filtering, so the image generator operates on a fully constrained prompt from
the outset.

\subsubsection{Review and Repair}

Each candidate image passes through a judging panel of three independent
judges. Every judge scores the image against seven criteria: \emph{relevance}
(alignment with the given scene), \emph{therapeutic value} (suitability for
open description tasks), \emph{story clarity} (legibility as a narrative
prompt), \emph{child-friendliness} (age-appropriate, safe content),
\emph{emotional expressiveness} (visible character emotions),
\emph{photorealism} (professional-grade visual quality), and \emph{cultural
    authenticity} (accurate representation of Middle Eastern setting and attire). All three
judges must independently pass the image, and a minimum score of 7.0 out of 10
is required on each criterion.

When an image fails, the judges' feedback is stored in short-term memory and
returned to the Prompt Generator as refinement feedback for the next attempt.
The repair loop runs for up to three attempts. If no attempt passes, the image
is discarded and the pipeline proceeds to the next scene in the batch.

\subsubsection{Memory Layers}

The pipeline maintains two distinct memory layers. \emph{Long-term memory}
stores accepted images together with their prompts, scene summaries, and
aggregate judge scores. It serves two roles simultaneously: the Prompt
Generator draws on recent successful examples as style references, and the
Director reads coverage context to know which settings, character types, and
themes have already been well represented. \emph{Short-term memory} is scoped
to a single generation cycle. It holds the feedback from the most recent failed
evaluation and is cleared when a new image cycle begins. Together, these two
layers connect the immediate repair loop with the longer-term goal of building
a diverse, non-redundant therapy library.

\begin{figure}[t]
    \centering
    \begin{minipage}[t]{0.32\textwidth}
        \centering
        \includegraphics[width=\linewidth]{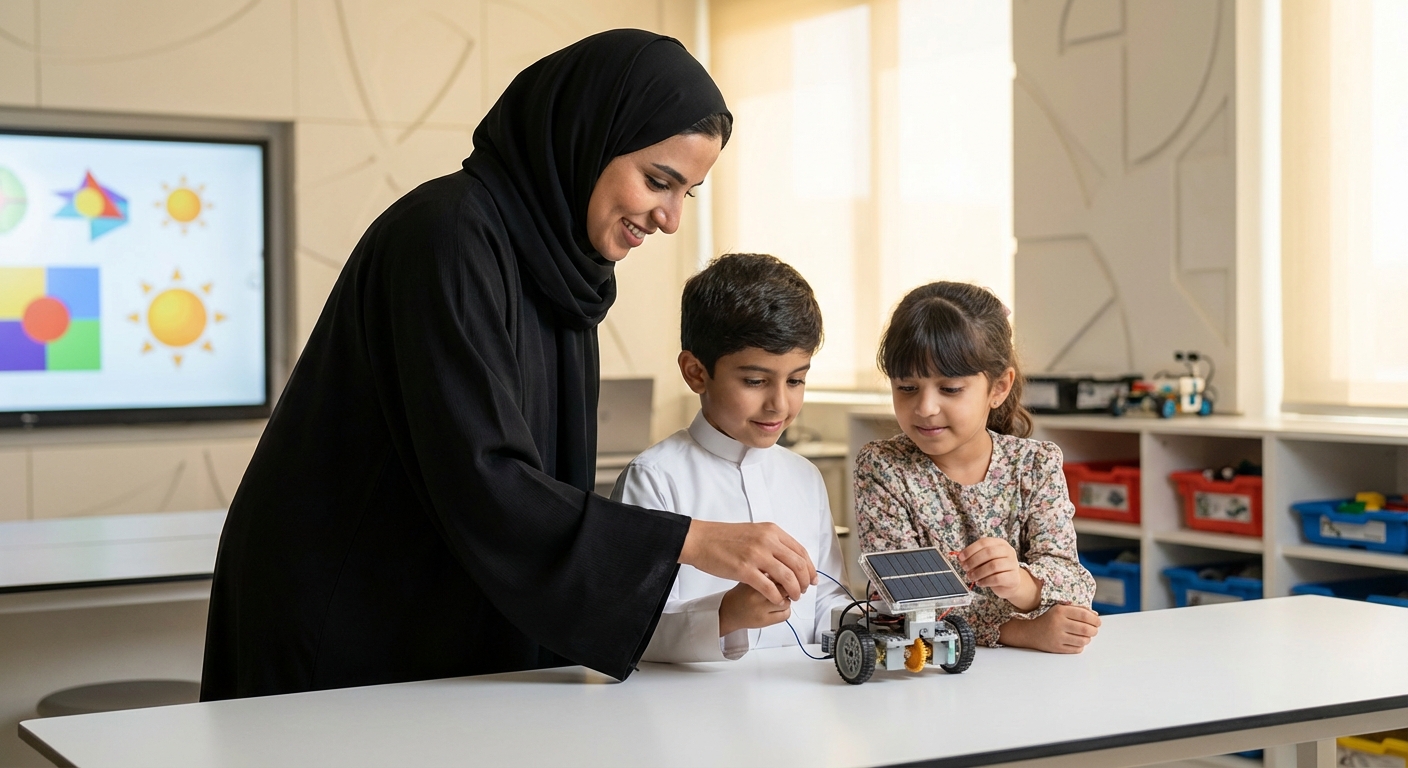}
    \end{minipage}\hfill
    \begin{minipage}[t]{0.32\textwidth}
        \centering
        \includegraphics[width=\linewidth]{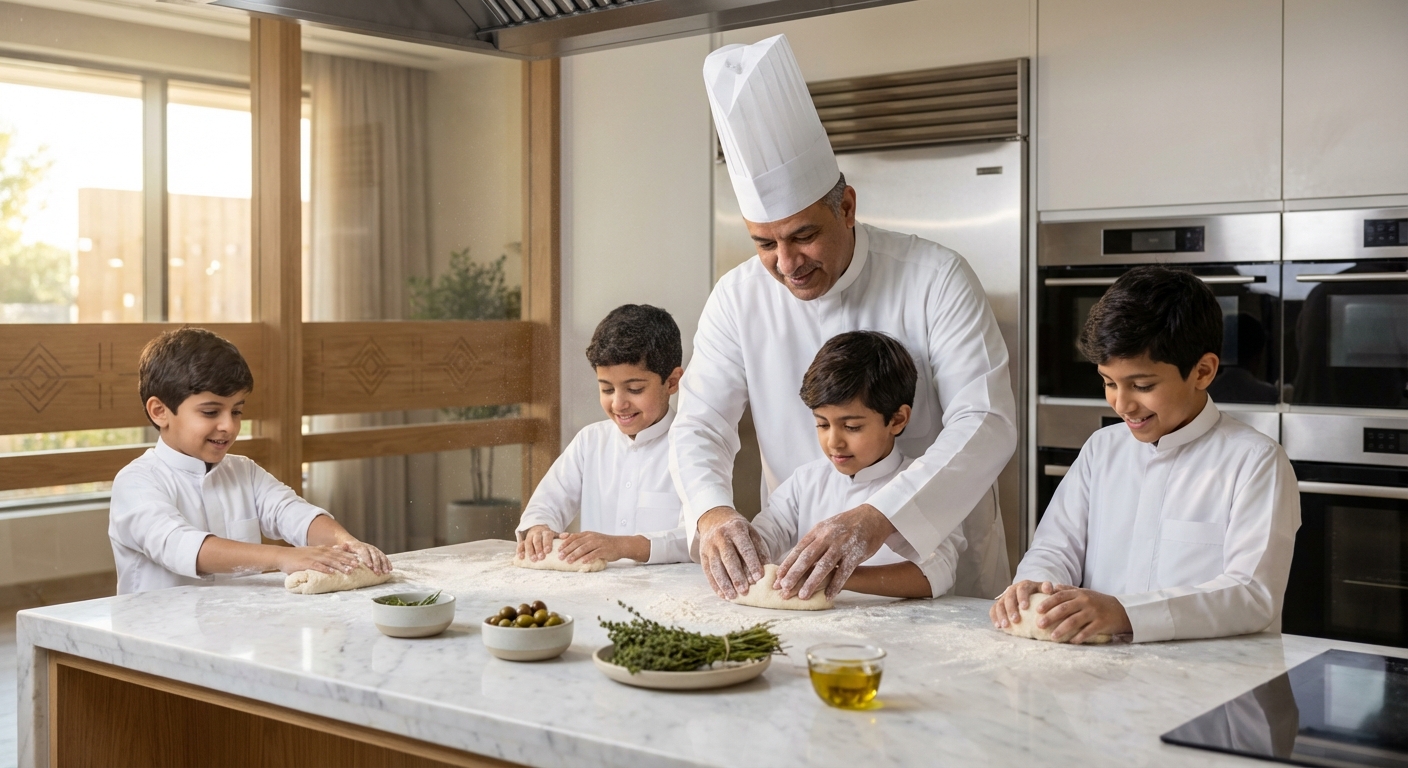}
    \end{minipage}\hfill
    \begin{minipage}[t]{0.32\textwidth}
        \centering
        \includegraphics[width=\linewidth]{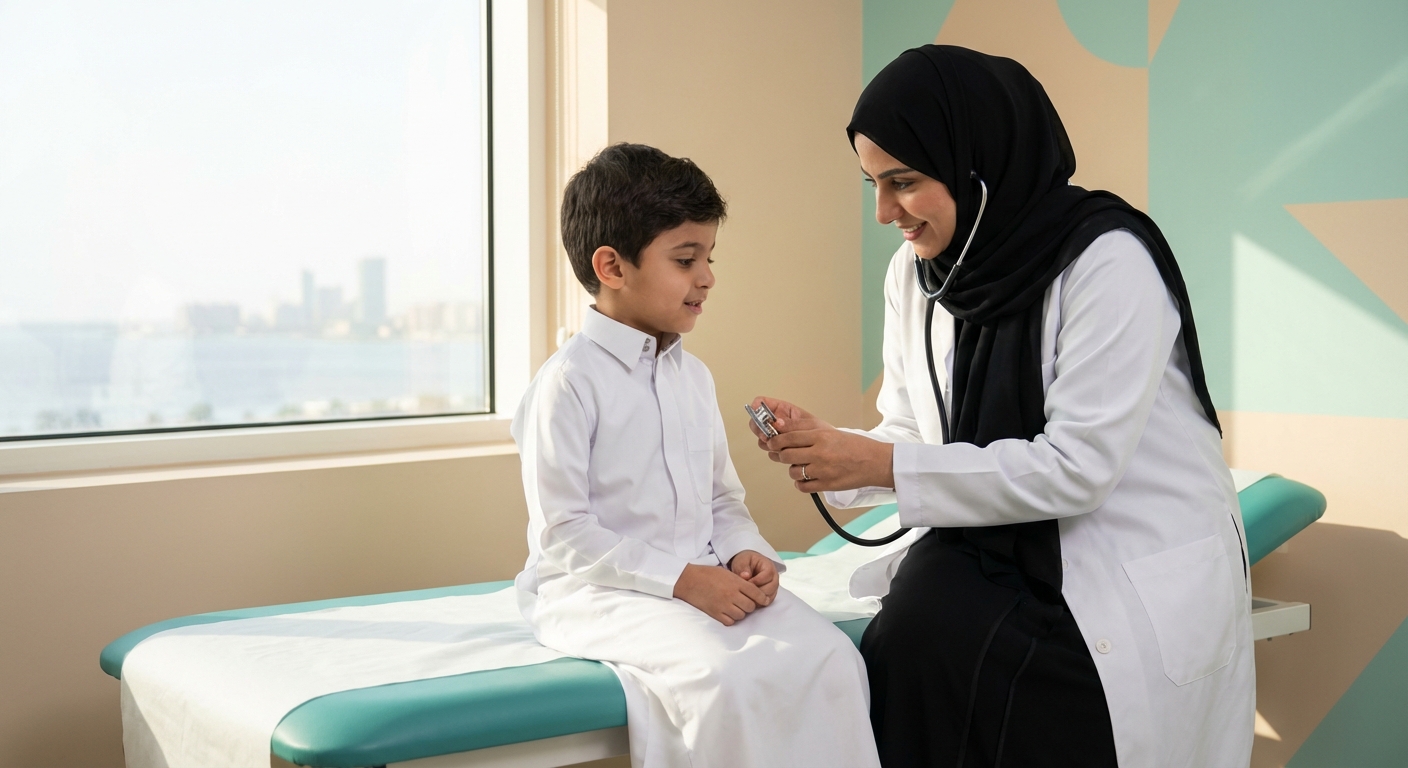}
    \end{minipage}
    \caption{Three example images from the ASDE-Picture Description pipeline: a classroom activity, a family cooking scene, and a pediatric clinic interaction. The scenes are culturally grounded in Middle Eastern settings and rich enough in characters, actions, and objects to support open-ended verbal description. 
    Some more examples are shown in Appendix Figure \ref{fig:appendix-study1-pd-asde}.}
    \Description{Three photorealistic images side by side: a woman in traditional attire working with children at a classroom table, adults and children preparing food in a kitchen, and a clinician examining a child in a medical setting. All depict Middle Eastern settings.}
    \label{fig:sde-pd-examples}
\end{figure}

Figure~\ref{fig:sde-pd-examples} shows three example outputs from this
pipeline. They illustrate the kind of range the system aims for: school-like
learning, everyday family activity, and clinical interaction. In each case, the
scene is intended to support open description rather than single-word naming.

\subsection{Language Therapy Pipeline}

\begin{figure*}[t]
    \centering
    \includegraphics[width=\textwidth]{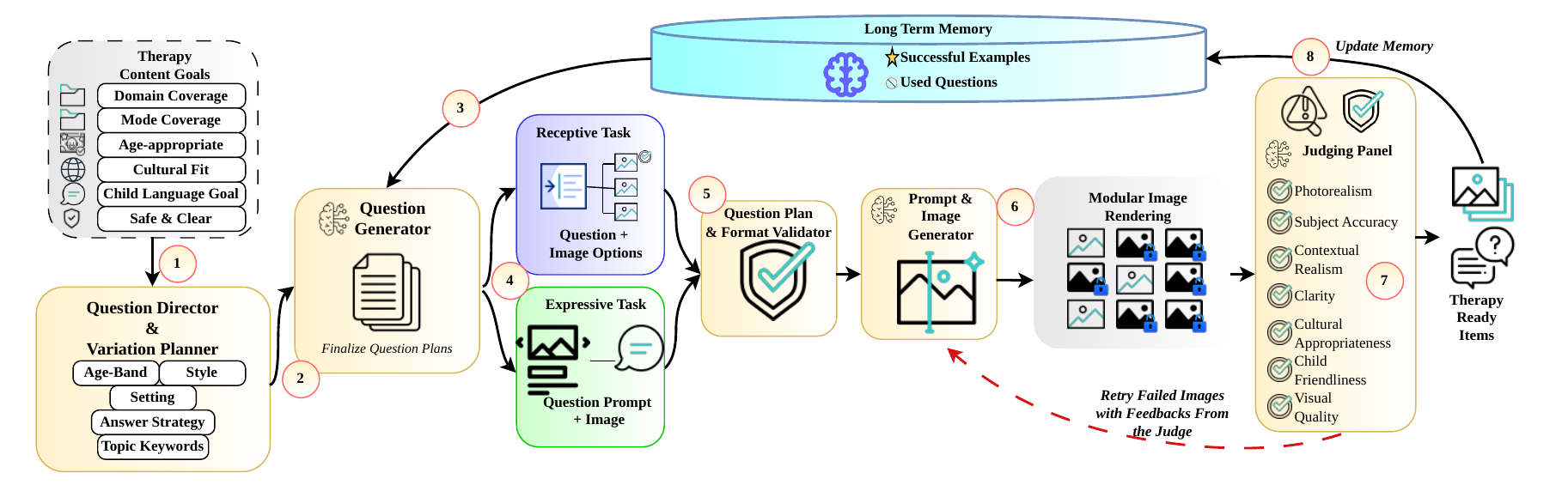}
    \caption{Architecture of the Agentic Synthetic Data Engine for Language Therapy (ASDE-LT). The pipeline generates structured question-and-answer therapy cards for vocabulary and language comprehension practice. The process begins with therapy content goals (left---domain coverage, mode coverage, age appropriateness, cultural fit, and safety requirements), which feed into a question director and variation planner that sets the degrees of variation for each generation run (e.g., age band, visual style, setting, answer strategy, and topic keywords). The question generator then produces candidate items using these planning signals together with long-term memory, which stores both successful question examples (as quality references) and previously used questions (to avoid repetition). Each generated item is then branched into one of two task formats: a Receptive task (multiple image-based answer options from which the child selects one) or an Expressive task (a single image prompt to which the child gives a spoken response). Both task types pass through a validator before image generation, allowing the system to catch poorly formed questions before investing in visual rendering. The image generator produces the required visual assets, and a judging panel evaluates them against criteria including photorealism, subject accuracy, contextual realism, clarity, cultural appropriateness, child friendliness, and visual quality. Accepted items become therapy-ready materials and update long-term memory; failed images are retried with judge feedback.}
    \Description{A pipeline diagram for language-therapy item generation. Therapy content goals on the left specify domain coverage, mode coverage, age appropriateness, cultural fit, child language goals, and safe and clear content. These feed a question director and variation planner that controls age band, style, setting, answer strategy, and topic keywords. The output goes to a question generator that also receives input from long-term memory containing successful examples and used questions. Generated items then branch into a receptive task with a question and image options, or an expressive task with a question prompt and image. Both paths pass into a validator, then into an image generator and a modular image rendering layer. A judging panel checks photorealism, subject accuracy, contextual realism, clarity, cultural appropriateness, child friendliness, and visual quality. Accepted items become therapy-ready items and update long-term memory, while failed images are retried using feedback from the judging panel.}
    \label{fig:sde-lt}
\end{figure*}

The language therapy pipeline solves a different problem. Here, the challenge
is not to create one rich scene, but to create structured question cards that
are clear, age-appropriate, and easy to score. The pipeline therefore needs
tighter control over the logic of each item. It also needs to support two
modes: receptive tasks, where the child chooses from image options, and
expressive tasks, where the child answers orally to a prompt or image. ASDE-LT
handles this through a set of interacting components that follow the flow shown
in Figure~\ref{fig:sde-lt}.

\subsubsection{Therapy-Goal and Variation Component}

The process begins with therapy content goals: domain coverage, mode coverage,
age appropriateness, cultural fit, child language goals, and the need for safe
and clear materials. These goals are passed to a question director and
variation planner, which sets the main degrees of variation for each run,
including age band, style, setting, answer strategy, and topic keywords. This
component is what keeps the pipeline tied to therapy needs rather than letting
it drift toward generic question generation.

\subsubsection{Memory-Informed Question Generation}

The question generator then uses those planning signals together with long-term
memory. In the figure, this memory stores successful examples and used
questions. Successful examples act as grounded references for what a good item
looks like, while used-question memory helps the system avoid repeating the
same concept or wording too often. This is especially important when the goal is
to build a reusable therapy library rather than a one-off batch.

\subsubsection{Task-Branching Component}

Once a candidate question is created, the pipeline branches it into the right
task form. For receptive mode, this means a question paired with multiple
image-only options. For expressive mode, it means a question prompt paired with
one image that supports a spoken response. This branching is not a cosmetic
difference. The two modes have different therapeutic roles and require
different visual structure, so they need to be handled explicitly before the
pipeline spends effort on rendering.

\subsubsection{Validation and Rendering Components}

Both task types then pass through a validator before image generation. This is
an important control point in the LT pipeline. It lets the system check whether
the question is coherent, answerable, and well formed before moving on to the
more expensive visual steps. After validation, the image generator produces the
needed assets, and the modular rendering layer assembles them into the final
set of card elements, such as multiple option images for receptive tasks or a
single supporting image for an expressive prompt.

\subsubsection{Judging and Memory-Update Component}

The judging panel reviews the rendered assets against the criteria shown in the
figure: photorealism, subject accuracy, contextual realism, clarity, cultural
appropriateness, child friendliness, and visual quality. If a rendered image
fails, the feedback is sent back to the image generator for targeted retry
rather than restarting the whole question from scratch; when an item passes, it
becomes a therapy-ready artifact and the system updates long-term memory,
adding new successful examples and recording the question as used. This closes
the loop between one generation run and the next.

\begin{figure}[t]
    \centering
    \begin{minipage}[t]{0.48\textwidth}
        \centering
        \includegraphics[width=\linewidth]{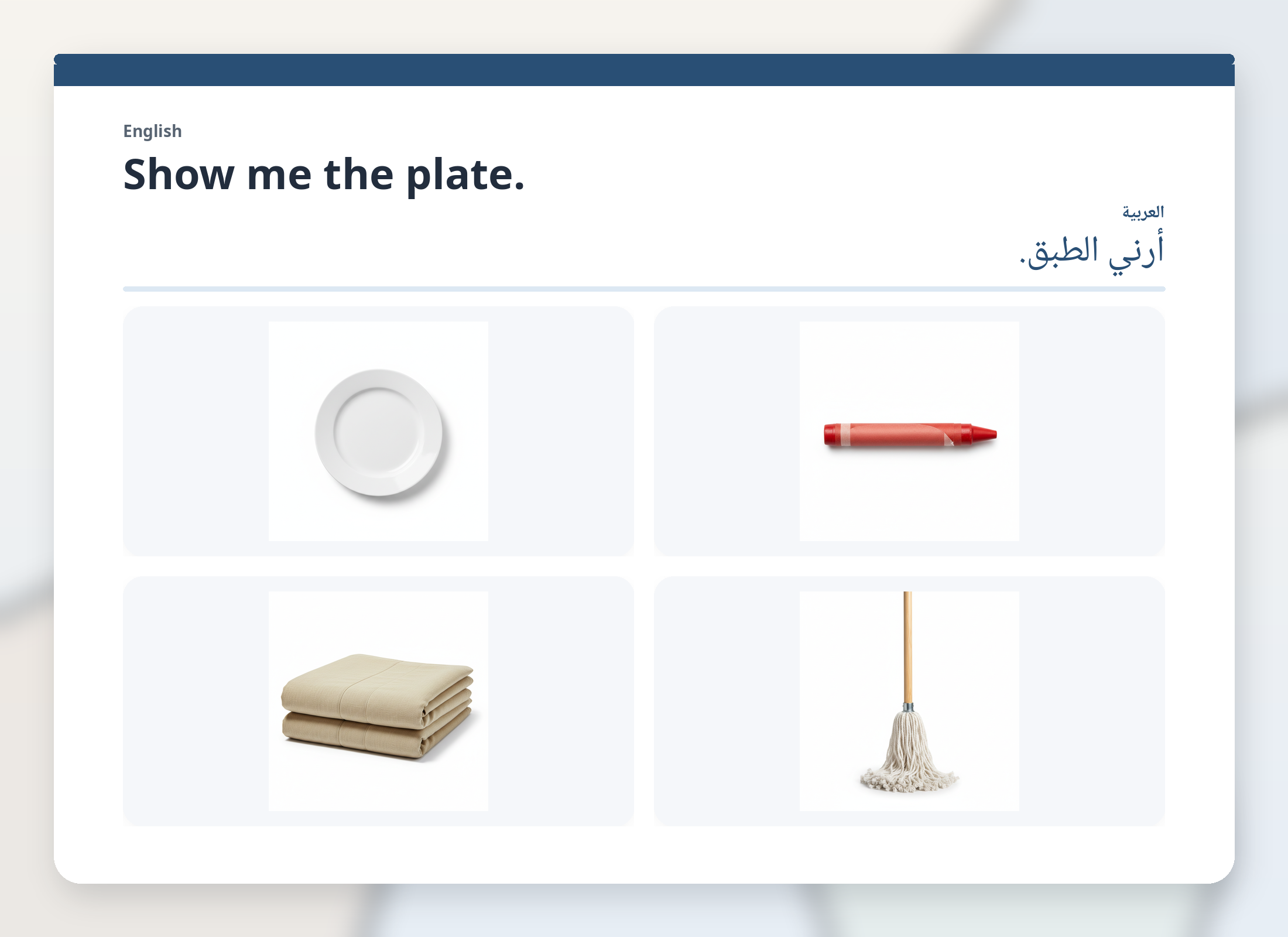}
    \end{minipage}\hfill
    \begin{minipage}[t]{0.48\textwidth}
        \centering
        \includegraphics[width=\linewidth]{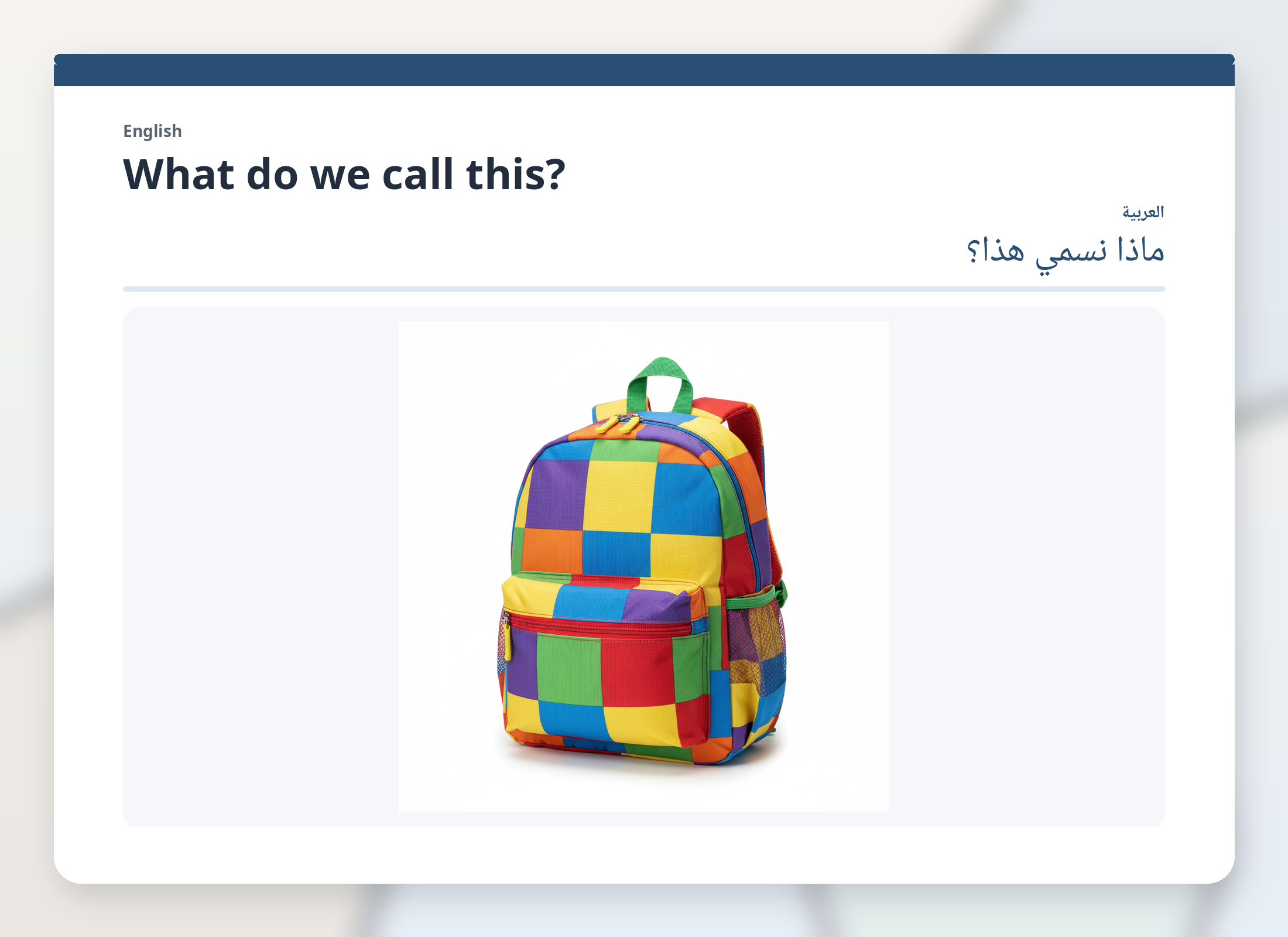}
    \end{minipage}
    \caption{Two example therapy cards from the ASDE-Language Therapy pipeline. Left: a Receptive card (``Show me the plate'' / Arabic: \textit{Arni al-tabaq}) with four image options. Right: an Expressive card (``What do we call this?'' / Arabic: \textit{Matha nusammi hadha}) with a single backpack image. The bilingual, image-led format supports Arabic-speaking children while maintaining English reference labels. 
    Some more examples are shown in Appendix Figure \ref{fig:appendix-study1-lt-asde}.}
    \Description{Two therapy cards side by side. The left card shows a Receptive task with a bilingual prompt and four image options. The right card shows an Expressive task with a prompt and a single backpack image.}
    \label{fig:sde-lt-examples}
\end{figure}

Figure~\ref{fig:sde-lt-examples} shows two example outputs from ASDE-LT: a
receptive card and an expressive card. These examples are anecdotal, but they
show the two main interaction styles the pipeline supports and the bilingual,
image-led format used in the platform.

\section{Expert Evaluation}\label{sec:evaluation}

We evaluated \textit{Digital Harf} through two linked expert studies. Study~1
tested whether raw ASDE-generated therapy items were clinically acceptable
without manual editing (90.1\% acceptance). Study~2 assessed the full platform
across five domains---clinical validity, cultural fit, usability,
personalization, and adoption---via Likert ratings and open-ended responses.
Key findings include strong cultural alignment (the highest-rated domain),
promising usability, and important signals about the need for evidence-based
validation before deployment.

\subsection{Participants, Procedure, and Response Structure}

Thirteen licensed SLPs contributed expert feedback: nine in Study~1, eleven in
Study~2, and seven in both. Experience ranged from 3 to 12 years across
hospital, rehabilitation, school-based, and center-based contexts in the
Middle East. Before data collection, participants attended a guided two-hour
walkthrough of the Arabic interface and inspected the modules hands-on.

Study 1 focused on item-level clinical acceptability. Experts reviewed 27
uncurated ASDE outputs sampled directly from the pipeline: 12 Picture
Description (PD) scenes and 15 Language Therapy (LT) cards. Study 2 focused on
the full system. Experts completed a structured questionnaire after reviewing
the platform, its therapeutic modules, and its expected home-use workflow.
Table~\ref{tab:qual-study-design} summarizes the two instruments.

\begin{table*}[t]
    \sffamily
    \small
    \def\arraystretch{1.05}\setlength{\tabcolsep}{0.28em}
    \centering
    \caption{Overview of the two expert evaluation studies. Study~1 assessed whether ASDE-generated therapy items were clinically acceptable without manual editing. Study~2 assessed the full platform across five domains. SLPs = licensed Speech-Language Pathologists; PD = Picture Description (scenes for verbal description); LT = Language Therapy (question-and-answer cards). Full questionnaire wording for Study-2 is in Table~\ref{tab:study2-domain-summary}. 
    Some of the Study~1 contents are reported in Appendix Section~\ref{sec:appendix-study1-asde-contents}, Figures \ref{fig:appendix-study1-pd-asde} and \ref{fig:appendix-study1-lt-asde}.}
    \label{tab:qual-study-design}
    \arrayrulecolor{appendixrule}
    \begin{tabularx}{\textwidth}{|L{0.10\textwidth}!{\color{appendixrule}\vrule}L{0.20\textwidth}!{\color{appendixrule}\vrule}C{0.11\textwidth}!{\color{appendixrule}\vrule}L{0.20\textwidth}!{\color{appendixrule}\vrule}Y|}
        \hline
        \rowcolor{appendixheader}\textbf{Study} & \textbf{Focus}                  & \textbf{Experts} & \textbf{Materials}                                             & \textbf{Response format}                                                                                  \\
        \hline
        Study 1                                 & ASDE content acceptability      & 9 SLPs           & 27 uncurated items: 12 PD scenes and 15 LT cards               & Binary acceptable or not acceptable decisions, followed by free-text comments                             \\
        \hline
        Study 2                                 & Full-platform expert assessment & 11 SLPs          & 16 Likert items across five domains, plus 6 open-ended prompts & 5-point Likert ratings and narrative responses about value, limitations, deployment, and caregiver burden \\
        \hline
    \end{tabularx}
    \arrayrulecolor{black}
\end{table*}

\subsection{Measures and Analytic Frame}

Study 1 used binary acceptability judgments because the central question was
practical: would a therapist use the item as-is with an Arabic-speaking child
with ASD? In Study 2, we used 5-point Likert ratings to capture expert
agreement with the platform's clinical fit, cultural grounding, interaction
design, usefulness, and likely uptake.

For readability, we use the following abbreviations throughout the results: CV
for \textit{clinical validity}, CLA for \textit{cultural and linguistic fit},
UID for \textit{usability and interaction}, POU for \textit{personalization and
    usefulness}, and PAI for \textit{adoption and impact}. Higher means indicate
stronger agreement, and we also report the share of responses at 4 or 5 to show
how consistently experts leaned positive rather than merely neutral. Open-ended
responses were coded inductively and then regrouped around recurring practical
concerns such as carryover beyond the clinic, Arabic material scarcity,
engagement, therapist trust, and caregiver support. Full questionnaire wording
is provided in Table~\ref{tab:study2-domain-summary}, and
detailed item-level statistics are reported in 
Appendix Section~\ref{sec:appendix-study1-asde-contents}, Table \ref{tab:appendix-study1-items}.

\begin{table*}[t]
    \sffamily
    \footnotesize
    \def\arraystretch{1.03}\setlength{\tabcolsep}{0.28em}
    \centering
    \caption{Study~2 questionnaire structure and expert ratings. The table organises 16 Likert-scale items (1 = strongly disagree, 5 = strongly agree) and six open-ended prompts across five evaluation domains, rated by 11 speech-language pathologists (SLPs). Shaded domain rows report the domain mean and the aggregate share of responses at 4 or 5 (\%~$\geq 4$); item rows report per-item mean, standard deviation (SD), and \%~$\geq 4$ (\%~$\geq 4$ is the share of responses at 4 or 5, showing how consistently experts leaned positive.); open-ended rows are unscored. Domain abbreviations: CV = Clinical Validity, CLA = Cultural and Linguistic Fit, UID = Usability and Interaction, POU = Personalization and Usefulness, PAI = Adoption and Impact, OEP = Open-ended Prompts.}
    \label{tab:study2-domain-summary}
    \arrayrulecolor{appendixrule}
    \begin{tabular}{|C{0.07\textwidth}!{\color{appendixrule}\vrule}L{0.47\textwidth}!{\color{appendixrule}\vrule}C{0.07\textwidth}!{\color{appendixrule}\vrule}C{0.07\textwidth}!{\color{appendixrule}\vrule}C{0.09\textwidth}|}
        \hline
        \rowcolor{appendixheader}\textbf{Code} & \textbf{Statement} & \textbf{Mean} & \textbf{SD} & \textbf{\%~$\geq 4$} \\
        \hline
        \rowcolor{appendixsubheader}
        \multicolumn{2}{|l!{\color{appendixrule}\vrule}}{\textbf{Clinical Validity (CV)}}
            & 3.80 & {---} & 68.2\% \\
        \hline
        CV1 & The system includes tasks that are consistent with established speech and language therapy practice.                          & 4.09 & 1.38 & 81.8\% \\
        CV2 & The modules are appropriate for supporting therapeutic goals in speech and language development.                             & 3.82 & 1.17 & 63.6\% \\
        CV3 & The system provides feedback that is relevant for clinical interpretation and decision-making.                               & 3.64 & 0.81 & 63.6\% \\
        CV4 & The platform can be used as a supplementary tool alongside therapist-led intervention.                                       & 3.64 & 1.43 & 63.6\% \\
        \hline
        \rowcolor{appendixsubheader}
        \multicolumn{2}{|l!{\color{appendixrule}\vrule}}{\textbf{Cultural and Linguistic Fit (CLA)}}
            & 4.15 & {---} & 78.8\% \\
        \hline
        CLA1 & The content is culturally appropriate for Arabic and Middle Eastern contexts.                                               & 4.27 & 0.79 & 81.8\% \\
        CLA2 & The materials reflect realistic and familiar environments for children.                                                     & 4.09 & 1.04 & 72.7\% \\
        CLA3 & The language used is appropriate for the target users.                                                                     & 4.09 & 0.94 & 81.8\% \\
        \hline
        \rowcolor{appendixsubheader}
        \multicolumn{2}{|l!{\color{appendixrule}\vrule}}{\textbf{Usability and Interaction (UID)}}
            & 3.97 & {---} & 78.8\% \\
        \hline
        UID1 & The system is easy to use and navigate.                                                                                    & 4.45 & 0.69 & 90.9\% \\
        UID2 & The interaction design is suitable for children with ASD in collaboration with parents or caregivers.                       & 3.45 & 1.04 & 63.6\% \\
        UID3 & The system can be used effectively in home settings.                                                                       & 4.00 & 0.89 & 81.8\% \\
        \hline
        \rowcolor{appendixsubheader}
        \multicolumn{2}{|l!{\color{appendixrule}\vrule}}{\textbf{Personalization and Usefulness (POU)}}
            & 3.55 & {---} & 57.6\% \\
        \hline
        POU1 & The system provides appropriately personalized feedback.                                                                   & 3.36 & 0.92 & 45.5\% \\
        POU2 & The system adapts well to different skill levels.                                                                         & 3.45 & 1.04 & 63.6\% \\
        POU3 & The system is useful for continuous practice outside therapy.                                                              & 3.82 & 1.40 & 63.6\% \\
        \hline
        \rowcolor{appendixsubheader}
        \multicolumn{2}{|l!{\color{appendixrule}\vrule}}{\textbf{Adoption and Impact (PAI)}}
            & 3.82 & {---} & 69.7\% \\
        \hline
        PAI1 & I would consider using this system in my practice.                                                                        & 3.91 & 1.38 & 72.7\% \\
        PAI2 & This system addresses a gap in current therapy resources.                                                                  & 3.64 & 1.29 & 63.6\% \\
        PAI3 & I would recommend that parents use this platform in home settings.                                                         & 3.91 & 1.38 & 72.7\% \\
        \hline
        \rowcolor{appendixsubheader}
        \multicolumn{5}{|l|}{\textbf{Open-ended
prompts (OEP)}} \\
        \hline
        OEP1  &  What are the main challenges you face when delivering speech and language therapy for children with Autism in your current practice?
                                                                                               & {---} & {---} & {---} \\
        OEP2  & What do you consider the most valuable aspect of this system for therapy?
                                                                                        & {---} & {---} & {---} \\
        OEP3  & What is the main limitation or concern you have with this system?
                                                                                         & {---} & {---} & {---} \\
        OEP4  & In what scenarios do you see this system being most useful?                                                                                         & {---} & {---} & {---} \\
        OEP5  & What improvements would you prioritize before large-scale deployment?                                                                                         & {---} & {---} & {---} \\
        OEP6  & Do you think this system can reduce caregiver burden? Why or why not?                                                                                        & {---} & {---} & {---} \\
        \hline
    \end{tabular}
    \arrayrulecolor{black}
\end{table*}

\subsection{Study 1: Clinical Acceptability of ASDE-Generated Content}

Study 1 was designed as a hard test of content readiness. We did not ask
therapists to evaluate hand-picked exemplars; we asked them to judge raw output
from the pipeline. By demonstrating strong acceptability rates under these
conditions, the ASDE clears a more meaningful bar than a curated showcase.

\begin{figure*}[t]
    \centering
    \includegraphics[width=\textwidth]{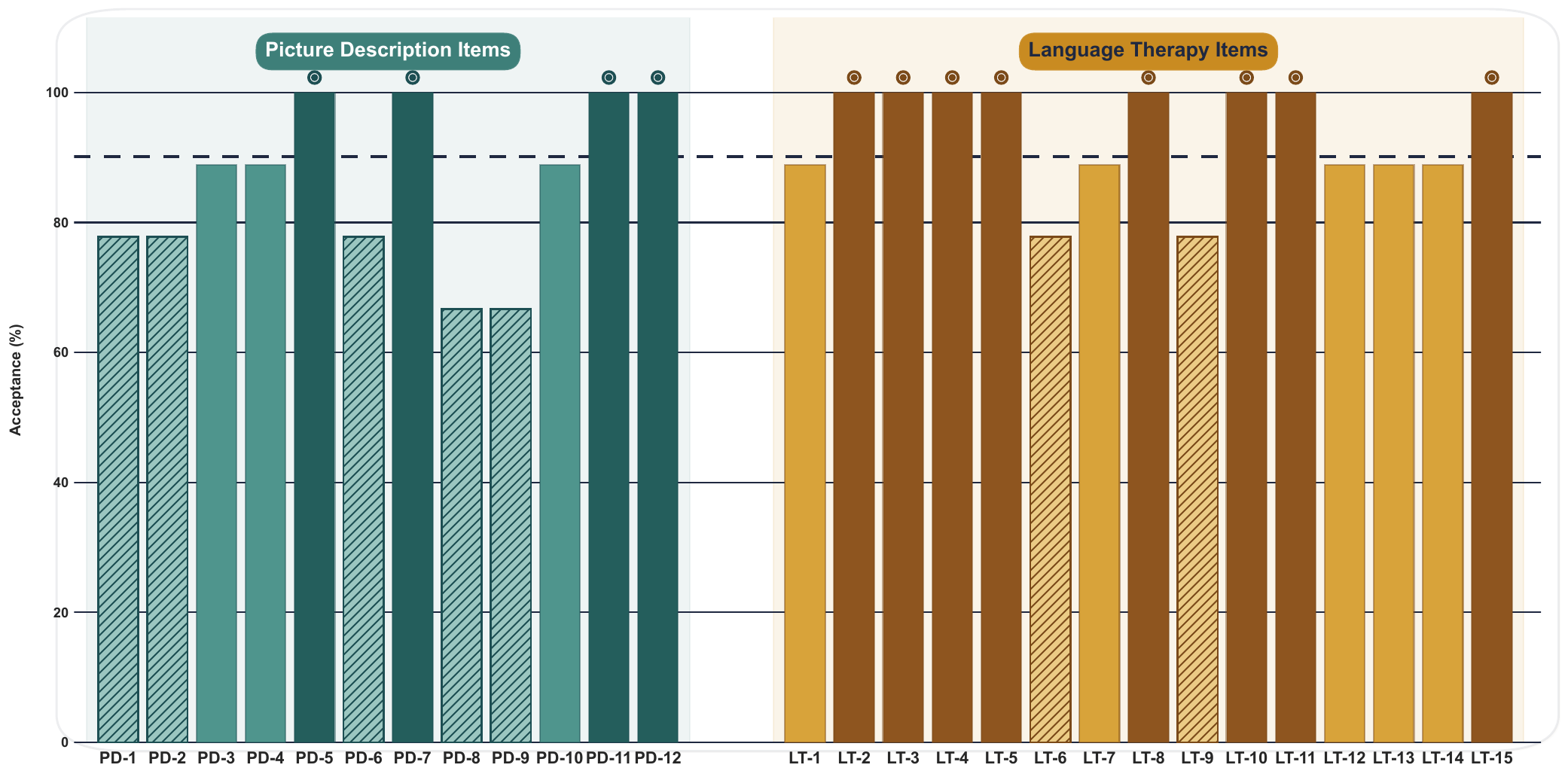}
    \caption{Per-item clinical acceptability rates for ASDE-generated content in Study 1. PD items showed the greatest spread, while LT items clustered more tightly near full acceptance. The dashed line marks the overall acceptance rate across all 243 expert-item judgments.}
    \label{fig:study1-acceptance}
    \Description{A wide bar chart of item acceptance percentages for Study 1. Twelve PD bars appear on the left and fifteen LT bars on the right. Most bars are high, with many reaching 100 percent. The lowest bars are PD-8 and PD-9 at roughly two thirds acceptance. A dashed horizontal line near 90 percent marks the overall acceptance rate, and small dots above selected bars indicate full-consensus items.}
\end{figure*}

\subsubsection{High Clinical Readiness of Generated Content}

Across 243 expert-item judgments (9 raters $\times$ 27 items), the ASDE content
was accepted 219 times, yielding an overall acceptance rate of 90.1\%. The
Language Therapy module performed especially well at 93.3\% (126 of 135
judgments), while Picture Description still reached a strong 86.1\% (93 of
108). The higher LT rate suggests that the structured, question-anchored format
of LT tasks aligns more naturally with clinical expectations than the
open-ended visual descriptions required by PD.

In the LT condition, 8 of the 15 items (53.3\%) received unanimous acceptance
from all 9 evaluators, while 4 of 12 PD items (33.3\%) achieved the same.
Across both modules, no item dropped below a two-thirds acceptance rate. That
pattern is important because the study was intentionally unforgiving. The items
were sampled from the pipeline without any post-hoc repair or curation. Under
those conditions, a 90.1\% acceptance rate indicates that the generation
process is not producing occasional good examples by chance; it is producing
therapy-ready content at a rate high enough to support clinical use.

\subsubsection{Impact of Scene Complexity on Picture Description}

The variation in Study 1 was concentrated in the PD module. While LT judgments
were relatively stable, PD exhibited greater variation, highlighting the
complexity of generating open-ended scenes. The two lowest-accepted items, PD-8
and PD-9, were each rejected by three independent raters, indicating consistent
rather than idiosyncratic disagreement. Items depicting common, single-action
scenes generally received higher acceptance (PD-5, PD-7, PD-11, PD-12), whereas
items featuring multiple competing scene elements or culturally specific but
low-familiarity activities tended to see lower acceptance.

This gradient is clinically meaningful for a system targeting children with
ASD, who are known to show heightened sensitivity to visual clutter and may
attend to salient peripheral elements rather than the intended target stimulus.
Reviewers did not describe the lower-scoring scenes as culturally inappropriate
in the abstract; they described them as harder to parse. Comments repeatedly
returned to split visual focus, cluttered backgrounds, or actions that were not
immediately readable. One senior evaluator with 11 years of experience
explicitly stated that ``it would help if each picture focused on a single,
clearly identifiable action.'' 

\subsubsection{Balancing Cultural Authenticity and Familiarity}\label{sec:balancing-cultural-authenticity-familiarity}

A recurring theme in evaluator commentary was the tension between cultural
authenticity and developmental familiarity. Experts broadly welcomed the
Arabic and Middle Eastern cultural grounding of the imagery. However, several
items that depicted hyper-specific cultural scenarios received lower acceptance
because children in clinical practice might not recognize them. One senior SLP
noted that images depicting rural occupational activities---honey extraction,
knitting, agriculturally-clothed figures---were culturally authentic but not
part of the everyday experience of the urban children typically seen in
clinical settings.

This feedback introduces a critical design distinction. Cultural authenticity
means that the content accurately represents the target culture. Cultural
familiarity means that the content reflects the actual lived environment of the
children using the system. The ASDE's current generation logic prioritizes
authenticity, but evaluator feedback suggests that familiarity should be
weighted more heavily, particularly for early-stage therapeutic content aimed
at children at lower developmental levels. As one evaluator recommended, target
actions should prioritize high-frequency, universally recognized verbs: ``eat,
drink, go, stop, sleep, wash, play, run, jump.'' We subsequently tuned the ASDE
judging criteria to effectively take this familiarity requirement into account.

Study 1 also exposed a smaller set of recurring but repairable issues that cut
across both modules: receptive LT prompts that could be solved through a noun
cue rather than the intended verb, semantically adjacent foils, image scale
problems, and bilingual phrasing that still needs review. These point to
concrete improvements in prompt design, visual QA, and linguistic checking
rather than to a failure of the overall approach.

\subsection{Study 2: Expert Assessment of the Full Digital Harf Platform}

\subsubsection{Overall Platform Assessment}

\begin{figure*}[t]
    \centering
    \includegraphics[width=0.92\textwidth]{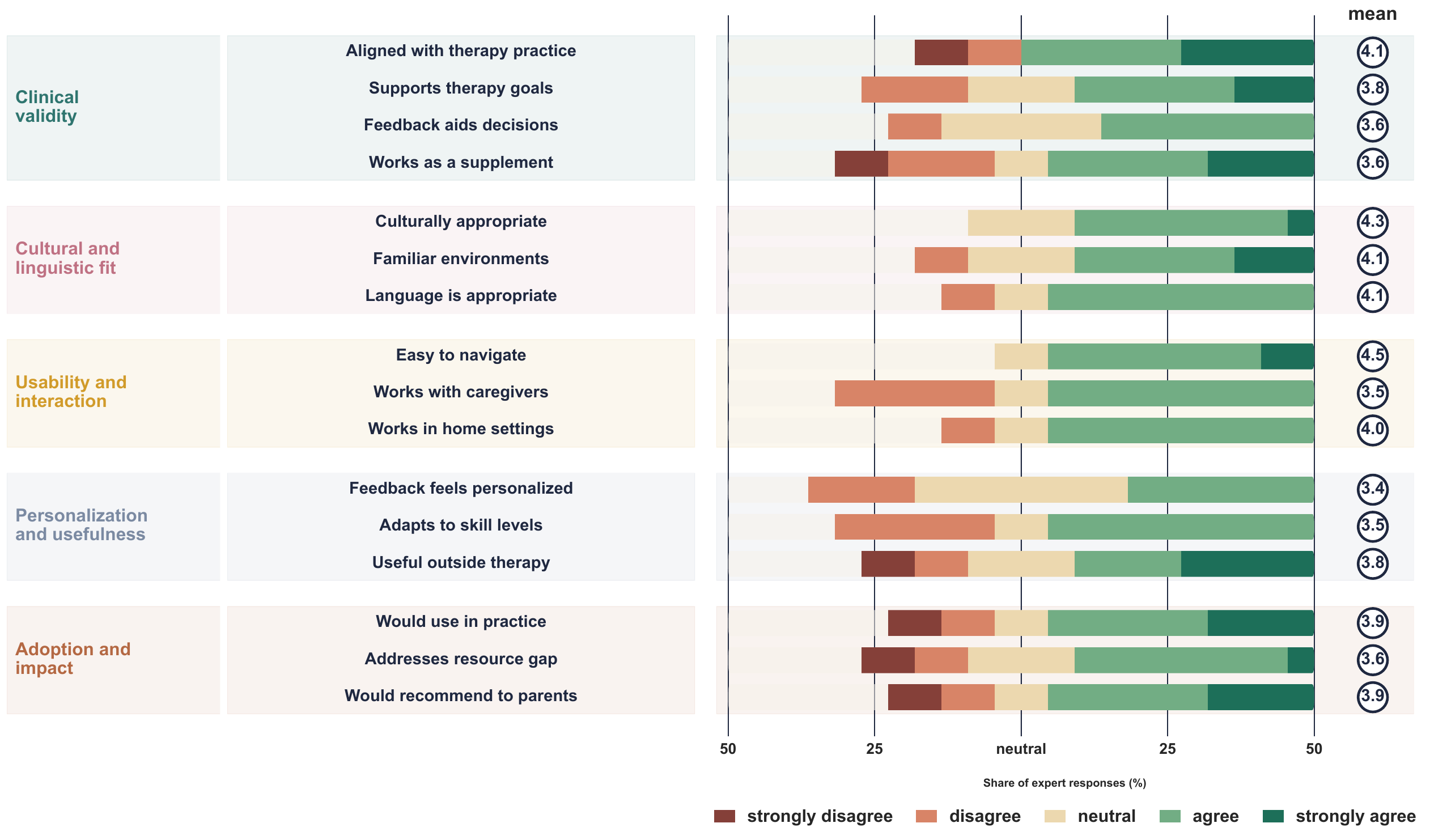}
    \caption{Distribution of Study~2 Likert-scale responses across all 16 items, grouped by evaluation domain. Each row shows one item rated by 11 SLPs (1 = strongly disagree, 5 = strongly agree). Diverging stacked bars show the spread across five response categories; circles report item means. Items are grouped into five domain blocks: Clinical Validity (CV), Cultural and Linguistic Fit (CLA), Usability and Interaction (UID), Personalization and Usefulness (POU), and Adoption and Impact (PAI). CLA shows the strongest positive concentration; POU shows the most spread toward neutral and negative responses.}
    \label{fig:study2-capsules}
    \Description{A diverging stacked-bar chart for Study 2 with sixteen rows grouped into five domain sections. CLA rows are concentrated on the agreement side; POU rows show more neutrality and disagreement.}
\end{figure*}

\begin{figure*}[t]
    \centering
    \includegraphics[width=0.92\textwidth]{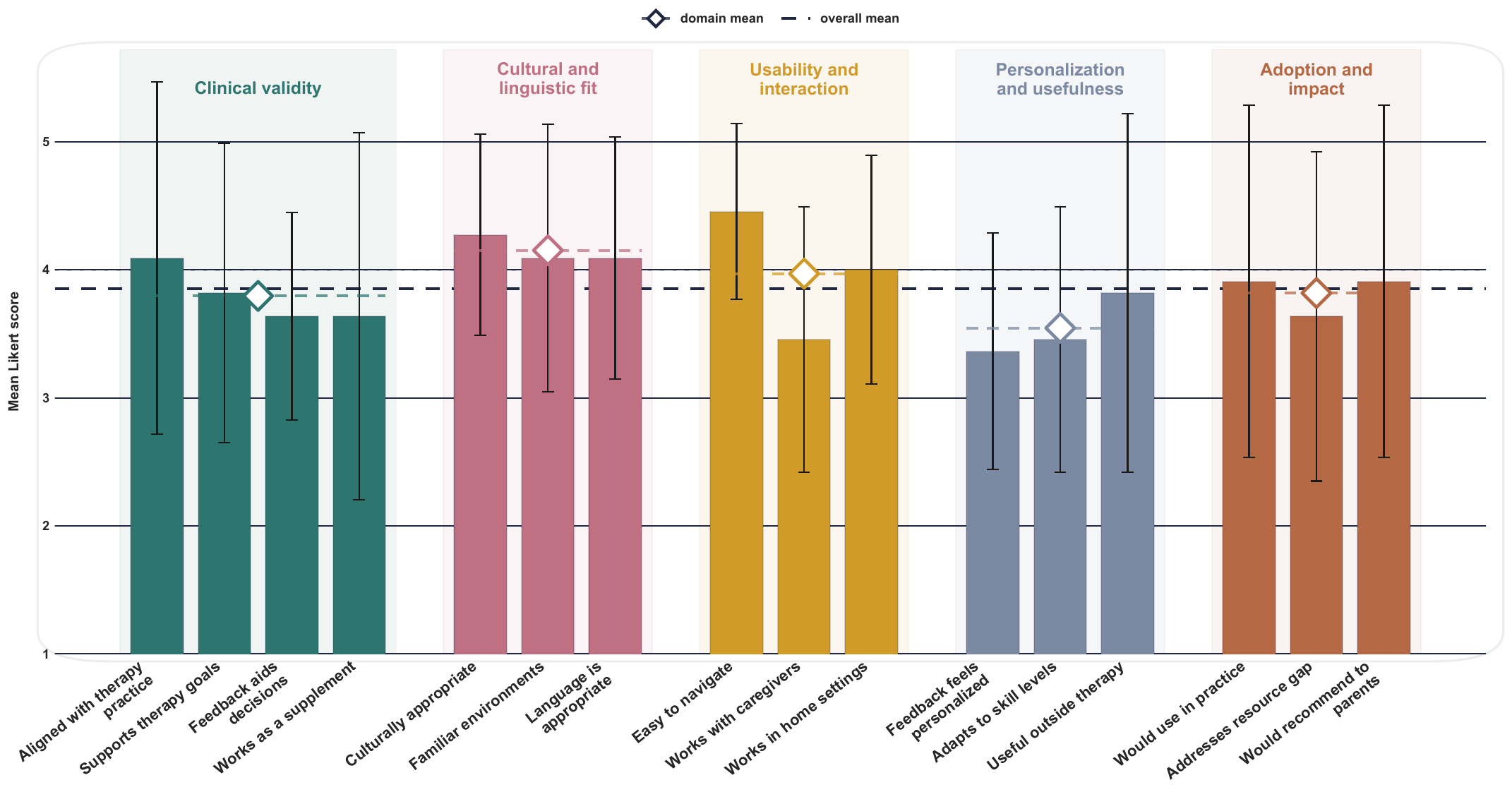}
    \caption{Study~2 item-level mean ratings with standard-deviation error bars, grouped by domain. Each bar is one of 16 Likert items rated by 11 SLPs. Diamond markers show domain means; the dashed line marks the overall system mean of 3.85. CLA items are all above 4.0 with tight error bars; POU items fall below 3.6 with wider spread, indicating both lower satisfaction and greater disagreement. This complements Figure~\ref{fig:study2-capsules} by showing magnitude and variability rather than response distributions.}
    \label{fig:study2-item-bars}
    \Description{A grouped bar chart of Study 2 item means with error bars, arranged in five domain groups. CLA bars exceed 4.0; POU bars fall below 3.6 with wider error bars. A dashed line at 3.85 marks the overall mean.}
\end{figure*}

Study 2 shifted from item readiness to system trust. The overall mean across
all 16 Likert items was 3.85, which placed the platform on the positive side of
neutral expert judgment. The shape of that positivity, however, was more useful
than the overall mean alone. At the domain level, CLA emerged as the strongest
area, while POU remained the weakest. At the item level, experts were most
confident that the platform was easy to navigate (UID-1, $M=4.45$) and less
confident that its personalization mechanisms were already mature (POU-1,
$M=3.36$; POU-2, $M=3.45$). Figures~\ref{fig:study2-capsules}
and~\ref{fig:study2-item-bars} show the same pattern from two complementary
views: distributional agreement and item-level means with variation.

\subsubsection{Cultural and Linguistic Fit}

At the domain level, Cultural and Linguistic Appropriateness (CLA) received the
strongest rating in the entire Study 2 instrument ($M=4.15$, with an average of
78.8\% of responses at 4 or 5). All three CLA items stayed above the 4.0
agreement threshold: cultural appropriateness ($M=4.27$, $SD=0.79$), familiar
environments ($M=4.09$, $SD=1.04$), and language appropriateness ($M=4.09$,
$SD=0.94$). The consistency across items is notable. CLA-1, measuring cultural
appropriateness for the Arabic and Middle Eastern context, was the
second-highest-rated item in the full instrument, and all three CLA items
achieved agreement rates between 72.7\% and 81.8\%, indicating coherent
satisfaction rather than isolated pockets of approval.

This finding supports the central argument for culturally specialized
platforms: cultural mismatch creates friction at the point of clinical use.
Study~2 provides practitioner-level evidence that the ASDE-driven content
pipeline, constrained by culturally specific prompting and Middle Eastern
visual and linguistic norms, produces materials that experts judge as
genuinely appropriate. One evaluator described the platform's core value as
``tools that help the specialist provide services appropriate to their
environment and language.''

Open-ended responses reinforced this pattern. Multiple evaluators independently
named Arabic-language content availability and cultural grounding as the
platform's most valuable aspect, describing cultural fit not as a surface
translation layer but as one of the platform's most credible contributions
given the persistent scarcity of appropriate Arabic therapy materials.

\subsubsection{The Need for Evidence-Based Validation}

The aggregate means in Study 2 conceal an important split in the response data.
The majority of evaluators (9 of 11) produced overall individual means between
3.38 and 4.75, indicating moderate to strong support across all five domains.
However, two respondents---one with 7 years of rehabilitation experience and
one with 12 years of hospital-based experience---rated the platform at an
overall mean of 2.31, substantially below the group average. Their skepticism
was tied explicitly to the platform's evidence base.

The evaluator with 12 years of experience stated: ``It is not evidence-based
approach. Development of evidence-based training content is needed.'' The
evaluator with 7 years of experience questioned whether a digital tool for
children with ASD could be endorsed without published clinical trial data. This
second evaluator's lower ratings extended beyond clinical validity into the CLA
domain as well ($M = 2.67$ across CLA items), suggesting that skepticism toward
the system's evidence base may spread into more critical appraisal across all
dimensions.

This polarization reveals a structural adoption gate: for digital health
tools, uptake depends not only on usability and cultural fit but on
\textit{evidence-based practice (EBP) criteria} that licensed clinicians are
trained to apply. As a novel AI-driven system without published clinical
outcome data, \textit{Digital Harf} faces a credibility challenge with EBP-oriented
practitioners regardless of technical quality. To address this, we plan to
conduct a comprehensive clinical trial as immediate future work to establish
evidence-based validation prior to deployment.

\subsubsection{Potential to Reduce Caregiver Burden}

One of the most consistent results in Study 2 was the agreement among
evaluators that \textit{Digital Harf} has meaningful potential to reduce caregiver
burden. Of the ten experts who provided a substantive answer to the
caregiver-burden question, seven said yes outright, two expressed conditional
support pending further refinement or training, and only one rejected the
claim.

The affirmative responses converged on three specific mechanisms. First,
\textit{material readiness}: the system removes the burden on parents to source
or improvise Arabic-language practice materials, which evaluators repeatedly
described as scarce in current practice. One evaluator noted that caregivers
``will have the goals ready and can access them easily.'' Second,
\textit{engagement transfer}: multiple evaluators cited children's natural
motivation with digital interfaces as a way to reduce the effort caregivers
must invest to initiate and sustain home practice sessions. One evaluator
articulated this directly: ``Children show notable attraction toward using
digital devices, which enhances their motivation to participate in therapeutic
activities.'' Third, \textit{structured continuity}: the system's session
format---pre-set goals, defined domains, automatic feedback---reduces the
cognitive and organizational burden on caregivers who lack SLP training. An
evaluator with 11 years of school-based experience described the system as
``reducing the effort in planning'' while ``involving the family in simple
ways.''

\subsubsection{Alignment with Practitioner Needs}

The clinical challenges practitioners cited most frequently in open-ended
responses mapped directly onto \textit{Digital Harf}'s capabilities.
Table~\ref{tab:challenge-mapping} presents this mapping. Generalization beyond
the clinic (4/9), inconsistent family follow-through (4/9), low engagement
(4/9), and Arabic material scarcity (3/9) each correspond to core
architectural decisions: home-based sessions with caregiver participation,
ASDE-driven content generation, and multimodal interaction with hint
scaffolding.

This convergence is not coincidental---\textit{Digital Harf} was designed with clinician
input---but its confirmation through independent evaluator testimony
strengthens the claim that the system addresses real clinical problems.
Practitioners who identified Arabic material scarcity as a top challenge
consistently named Arabic language availability as the platform's most valuable
aspect, showing that the platform solves a problem practitioners already feel.

\begin{table*}[t]
    \sffamily
    \small
    \def\arraystretch{1.05}\setlength{\tabcolsep}{0.28em}
    \centering
    \caption{Cross-study mapping from practitioner challenges to \textit{Digital Harf} capabilities. Frequencies (e.g., ``4/9'') indicate how many of nine experts raised each challenge. Cross-study evidence draws on both Study~1 and Study~2. The convergence between independently reported challenges and platform capabilities provides evidence that the system addresses real clinical problems rather than hypothetical ones.}
    \label{tab:challenge-mapping}
    \arrayrulecolor{appendixrule}
    \begin{tabularx}{\textwidth}{|L{0.18\textwidth}!{\color{appendixrule}\vrule}Y!{\color{appendixrule}\vrule}L{0.20\textwidth}!{\color{appendixrule}\vrule}L{0.22\textwidth}|}
        \hline
        \rowcolor{appendixheader}\textbf{Clinical challenge}                  & \textbf{Cross-study evidence}                                                                                                                       & \textbf{Mapped platform capability}                               & \textbf{Why the mapping matters}                                                                                 \\
        \hline
        \cellcolor{appendixsubheader}Generalization beyond the clinic (4/9)   & Study 2 responses repeatedly pointed to carryover difficulty; Study 1 comments favored clearer, everyday stimuli that can travel into home practice & Home-based LT and PD sessions with guided caregiver participation & Shows that experts linked content clarity to real carryover work rather than to isolated in-clinic performance   \\
        \hline
        \cellcolor{appendixsubheader}Arabic-material scarcity (3/9)           & Experts described a shortage of culturally appropriate Arabic resources; CLA was the highest-rated domain in Study 2                                & ASDE-driven Arabic and culturally constrained content generation  & Confirms that the platform's most valued strength addresses a live resource gap                                  \\
        \hline
        \cellcolor{appendixsubheader}Low engagement in structured tasks (4/9) & Reviewers asked for more interactive and game-like presentation; weaker ratings clustered around caregiver-facing interaction and personalization   & Multimodal interaction, hints, and child-facing activity flow     & Points to engagement as the next design pressure after content quality and cultural fit                          \\
        \hline
        \cellcolor{appendixsubheader}Inconsistent family follow-through (4/9) & Caregiver-burden responses emphasized ready-made goals, easier access, and more structured continuity                                               & Guided home workflow and therapist-to-caregiver continuity        & Explains why burden reduction was one of the most convergent positive findings                                   \\
        \hline
        \cellcolor{appendixsubheader}Lack of standardized Arabic tools (2/9)  & Experts named the absence of standardized, culturally appropriate resources; Study 1 showed high acceptability for raw generated items              & Reusable LT and PD materials with interpretable scoring support   & Suggests that the platform can function not only as practice software, but also as structured assessment support \\
        \hline
    \end{tabularx}
    \arrayrulecolor{black}
\end{table*}

\section{Discussion}\label{sec:discussion}

We organize the discussion around six themes. The first three address what
the evaluation revealed about the platform itself: cultural content as
infrastructure, clinical trust and the evidence-based practice gate, and the
caregiver as a design stakeholder. The fourth presents early caregiver signals
from the ongoing clinical validation study. These lead into a broader argument
situating the platform within the structural realities of the Arab world, made
concrete through KSA as a representative case. We close with limitations and
future directions.

\subsection{Cultural Content as Therapeutic Infrastructure}

A central finding is that cultural and linguistic grounding is not a surface
feature---it is infrastructure. CLA was the highest-rated domain in Study~2
($M=4.15$, 78.8\% at 4 or 5), and evaluators independently named
Arabic-language content availability as the platform's most valuable aspect.
This was not a reaction to cosmetic localization; experts described cultural
fit as credible precisely because they experience the absence of appropriate
Arabic materials as a daily constraint.

Much of the existing literature treats localization as a downstream step: build
the system first, translate later. Our experience suggests the opposite. When
therapeutic materials are not culturally native, the system fails at the point
of use regardless of algorithmic quality. The content gap is not a translation
problem; it is an infrastructure problem that requires purpose-built generation
pipelines. The ASDE's 90.1\% acceptance rate for uncurated content provides
early evidence that such pipelines are feasible.

The evaluation also revealed a useful nuance: the distinction between cultural
\textit{authenticity} (content that accurately represents the target culture)
and cultural \textit{familiarity} (content reflecting children's actual lived
experience). Several items that were culturally authentic received lower
acceptance because they depicted scenarios outside the everyday experience of
urban children typically seen in clinical settings. This distinction suggests
that familiarity should be weighted more heavily than authenticity in
early-stage therapeutic content.

\subsection{Clinical Trust and the Evidence-Based Practice Gate}

Clinical adoption depends on more than usability---it depends on
trust~\cite{suh2024opportunities}. While 9 of 11 evaluators rated the platform
positively, two experienced practitioners rated it substantially lower,
explicitly citing the absence of published clinical outcome data. This
polarization reflects a structural adoption gate: licensed clinicians are
trained to apply evidence-based practice (EBP) criteria, and digital tools
without outcome data face credibility challenges regardless of technical
quality.

For \textit{Digital Harf}, this means the expert evaluation reported here is necessary
but not sufficient. Recognizing this, the platform is now entering a
structured clinical validation study across multiple therapy centers, hospitals,
and medical institutions in the region, with participant enrollment underway.
The skeptical evaluators' responses also revealed a subtler dynamic:
they were not merely requesting data but protecting professional standards by
which specialist judgment coexists with automation. Systems like \textit{Digital Harf}
must position themselves as clinician-supporting rather than
clinician-replacing, building evidence incrementally rather than claiming
validation prematurely.

\subsection{The Caregiver as a Design Stakeholder}

\textit{Digital Harf} extends therapy into the home, where caregivers become the primary
facilitators. Seven of ten experts who responded to the caregiver-burden
question affirmed the platform's potential to reduce burden on families,
converging on three mechanisms: material readiness, engagement transfer through
digital interfaces, and structured continuity through pre-set goals and
automatic feedback. These findings align with HCI work positioning caregivers
as active participants rather than passive
recipients~\cite{dangol2025iwantslp,choi2025aacess}.

The caregiver dimension intersects with the content problem in a way that is
easy to underestimate. When therapy materials are culturally misaligned, the
burden of ``making it work'' falls disproportionately on caregivers. By
generating culturally native materials from the start, ASDE shifts this burden
from the family back to the system---a design choice with direct implications
for equity and access.

\subsection{Early Caregiver Signals}

The expert evaluations establish professional credibility but do not capture
the voice of the families who use the platform daily. \textit{Digital Harf} has now
entered a clinical validation study spanning multiple therapy centers, hospitals,
and medical institutions across the region. While full outcome reporting is
reserved for a dedicated future paper, early questionnaire responses have
already begun arriving from one of the participating sites, with enrollment
actively ongoing there and across the other sites. Among the first responses
received, five parents have completed the initial intake survey thus far. These
responses were not part of a planned interim analysis; they represent the
leading edge of a still-growing dataset and should be read as early directional
signals rather than validated findings.

One parent responded to the open-ended prompts in detail. The response was
specific and grounded: on what they valued most, the parent noted that the
platform ``makes it easier for families to organize daily activities''---directly
corroborating the structured-continuity benefit that SLP evaluators also
highlighted. On challenges, the parent flagged difficulty in
``interpreting results,'' surfacing a feedback transparency gap that did not
emerge in expert review. On improvements, they suggested ``adding
encouragement symbols and sounds,'' a concrete gamification direction
consistent with engagement principles for children with ASD. These early
signals triangulate with the SLP findings, surface a caregiver-specific
usability gap, and point to a design iteration the team will act on ahead of
the full cohort study.

\subsection{Why the Arab World Needs This Now: A Regional Imperative}

The thematic findings above speak to clinical and design validity. This final
analytical subsection situates those findings within a structural argument: the
Arab world faces a confluence of scale, scarcity, and cultural specificity
that makes pervasive, AI-driven therapy not just useful but arguably the only
viable path to equitable care.

Arabic-speaking populations are distributed across more than 20 countries with
a combined population exceeding 400 million, yet they are profoundly
underrepresented in the digital health literature~\cite{almurashi2022asdreview,lewis2025cld}.
Autism prevalence estimates across the Arab world vary widely---from below 0.1\%
to above 2\%---not because the underlying rates differ dramatically, but because
standardized surveillance infrastructure is largely absent~\cite{khamees2025estimating}.
This data gap itself signals the depth of the resource problem: systems for
counting affected children are still being built, while hundreds of thousands
of those children are already in need of intervention.

\paragraph{KSA as a Representative Case.}
To make this structural gap concrete, we examine the Kingdom of Saudi Arabia
(KSA)---a country that is, by regional standards, among the most resourced
and data-rich, making it a conservative lower bound on the severity of the
problem across the broader Middle East. KSA has one of the GCC's most
developed disability registries and a significant national investment in autism
services, and yet the gap remains stark.

Epidemiological studies place the prevalence of ASD among Saudi children at
1.33\% nationally, rising to 2.50\% in urban-concentrated studies such as
those from Riyadh~\cite{khamees2025estimating}. Translating even the
conservative national figure to an absolute count, the estimated number of
Saudi children with ASD ranges from approximately 37,000 to
202,000~\cite{khamees2025estimating}. The King Salman Center for Disability
Research estimates the total number of individuals with ASD in the Kingdom at
approximately 100,000~\cite{kscdr2024stats}. The annual cost of caring for a
single autistic child has been estimated at approximately 100,000
SAR ($\sim$\$27,000 USD)~\cite{tinelli2023economic}, placing the aggregate \textbf{yearly economic
burden of autism in KSA at approximately 10 billion SAR ($\sim$\$2.7 billion USD)}~\cite{kscdr2024stats,tinelli2023economic}---a
figure that does not include the incalculable indirect costs borne by families
who restructure their employment and daily lives around care obligations.

Against this scale, service capacity falls drastically short. Despite significant
national investments, there remains a severe structural gap between the growing
demand for autism services and the availability of specialized provision. Current
clinical volumes address only a small fraction of the estimated need. Crucially,
this shortage is not randomly distributed: existing services are heavily
concentrated in major urban centers, while families in peri-urban and rural
areas face a near-total absence of specialist access.

\paragraph{Generalizing Across the Region.}
If the patterns observed in KSA hold even approximately across the region---a
reasonable hypothesis given the shared linguistic context and the relative
wealth of KSA compared to many Arab states---the situation in less-resourced
countries is likely substantially worse. Countries across the Levant, North
Africa, and the broader Gulf share the same linguistic and cultural context as
this platform's target users, yet possess a fraction of the specialist
workforce per capita~\cite{attwell2022ostreview}. The content gap \textit{Digital Harf}
addresses is therefore not a KSA-specific problem; it is an Arabic-language
problem, a Middle East problem, and ultimately a problem of scale that no
workforce-dependent solution can solve alone.

This is precisely the domain where AI-driven pervasive computing holds its
greatest unrealized promise. By embedding structured, SLP-informed therapy
into daily home routines, a platform like \textit{Digital Harf} multiplies the reach
of each trained clinician without requiring proportional growth in the
professional workforce. The economic stakes are substantial: at 10 billion SAR
in annual burden for a single country, interventions that reduce even a
fraction of per-child care intensity could yield meaningful societal
returns---though demonstrating such effects requires the longitudinal outcome
studies discussed in our limitations. \textit{Digital Harf} does not solve the Arab
world's autism care crisis. What it offers, rather, is early evidence that
the technical and design foundations for closing this gap are emerging, and a
concrete instantiation that the field can build on.

\subsection{Limitations and Future Work}

The evaluation reported here is an expert assessment, not a full clinical trial.
While the SLP evaluations provide evidence of feasibility, clinical relevance,
and cultural alignment, they do not measure child-level outcomes such as
language gains, engagement over time, or behavioral change. A structured
clinical validation study is now underway across multiple therapy centers,
hospitals, and medical institutions in the region, with interaction logging
infrastructure in place for fine-grained collection of speech samples, response
patterns, and longitudinal progress data. Full cohort outcomes will be reported
in a dedicated future paper.

Additionally, the system's therapeutic modules currently focus on language
therapy, speech intelligibility, and picture description. Expanding to
additional modalities---such as pragmatic language, social communication
scenarios, or phonological awareness---would broaden clinical utility and is a
natural direction for future development in continued collaboration with
clinical partners.


The platform
was developed and evaluated primarily within Gulf Arabic contexts; Arabic
encompasses a wide spectrum of dialects---Gulf, Egyptian, Levantine, Maghrebi,
and others---that differ substantially in phonology, vocabulary, and everyday
register. Content and evaluation instruments calibrated to one dialect may not
transfer cleanly to others, and extending coverage across this dialectal
diversity is an important direction for future work.

\section{Replication Blueprint for Low-Resource Settings}\label{sec:framework}

A central design commitment of \textit{Digital Harf} is that clinical and cultural adaptation should happen at the content layer, not the algorithmic layer. This section makes that separation concrete, providing a practical guide for teams adapting the system to new languages and cultural settings. We separate what is reusable without
modification (the therapeutic interaction logic, personalization algorithms, and
pipeline architecture) from what must be rebuilt locally (cultural guidelines,
clinician-authored exemplars, and language-specific configuration). We close
with a recommended sequencing for resource-constrained teams.

\textit{Digital Harf} was built for Arabic in a Middle Eastern context, but the parts
that would need to change for a new language or culture are deliberately
separated from the parts that stay the same. The therapeutic interaction loop---structured tasks,
adaptive content selection, AI-scored feedback, and caregiver-facing
summaries---is language-agnostic. What is language- and culture-specific is the
content that fills these structures: which words children practice, what images
they see, how prompts are worded, and what counts as culturally appropriate.

\subsection{What Stays the Same}

The following components are reusable without modification across languages:

\begin{itemize}[nosep]
    \item The task interaction formats (receptive multiple-choice, expressive
          audio response, picture description with multi-attempt scoring, and
          interactive visual storytelling).
    \item The personalization algorithms. The picture description module's
          pool-based adaptive image selection works on performance scores, not
          language. The speech intelligibility module's AI-driven item generation
          analyzes historical accuracy and pronunciation patterns to personalize
          difficulty, and this logic transfers to any language for which a
           a sufficiently accurate speech-to-text service is available.
    \item The caregiver assistant's architecture: it retrieves session data
          through structured tool calls and generates progress summaries. The
          tools themselves query module-agnostic tables (scores, timestamps,
          attempt counts), so the assistant's reasoning layer carries over unchanged; the main adaptations needed are its system prompt and output language, though summary quality may vary across languages depending on the underlying model's coverage.
    \item The ASDE's pipeline structure: plan-then-render generation, multi-judge
          quality gates, long-term memory for variation, and
          stage-level concurrency control. These remain
          identical regardless of language.
\end{itemize}

\subsection{What Needs to Be Rebuilt Locally}

\paragraph{Cultural Guideline Sheets.}
Both SDE pipelines embed an explicit
\emph{cultural guideline document} that constrains every generated image. In
\textit{Digital Harf}, this document specifies Middle Eastern attire norms (e.g., abaya and
hijab for adult women, thobe and shemagh for men), acceptable social
interactions, environment details (Najdi architecture, date palms, desert
landscapes), and content prohibitions. For a new setting, this is the single most
important artifact to author. It should be written by someone who understands
both the local culture and child therapy contexts, because vague or generic
guidelines produce images that feel wrong to local clinicians and families. The
system enforces this document at every image generation step, so getting it
right has outsized impact.

\paragraph{Clinician-Authored Exemplar Bank.}
The language therapy pipeline does not generate questions from scratch. It
learns from a bank of clinician-written sample questions that anchor wording
style, answer structure, and difficulty. \textit{Digital Harf}'s bank covers 12
subdomains (common objects, actions, colors, shapes, categories, opposites,
negatives, associations, prepositions, yes/no questions, parts/whole, and WH
questions) with a handful of examples each~\cite{PAELS_TPT}. In a new language, the critical
first step is to have a speech-language pathologist write even a small set of
representative questions per subdomain. These examples teach the generation
model what a good question looks like in the target language, which matters far
more than increasing the generation budget.

\paragraph{Subdomain and Variation Configuration.}
Each subdomain carries its own topic lists (e.g., \emph{common fruits},
\emph{classroom items}, \emph{ball and box scenes}), prompt style templates,
and answer strategies (e.g., ``use one familiar everyday object with one obvious
name''). These are all defined in a single human-editable configuration file. The variation layer also rotates age bands, settings (home, school,
park, clinic), and topic keywords to keep generated content diverse. Localizing
these lists requires clinical input but no code changes.

\paragraph{Translation and Language Layer.}
\textit{Digital Harf} leverages a translation layer that converts generated English-language therapy questions into Arabic while preserving clinical semantics. This step can be performed by any large language model proficient in both English and the target language. Extending to a new language simply involves directing this process toward the desired language. The caregiver assistant’s system prompt should also be adapted into the local language with culturally appropriate tone and phrasing. UI text and TTS voice selection similarly require localization. For Arabic-script and other RTL languages, these introduce non-trivial rendering and voice-selection considerations beyond standard internationalization.

We adopted this two-stage design to leverage the larger pool of English-language clinical exemplars and stronger generation quality in English, at the cost of a translation step that must be clinically validated. Generating directly in the target language is also supported by the architecture and may better preserve idiomaticity and register; either path is viable.

\subsection{Practical Sequencing}

For teams with limited resources, we suggest the following order:

\begin{enumerate}[nosep]
    \item \textbf{Write the cultural guideline document and the sample question
              bank.} These are the two artifacts that most determine whether
          generated content feels clinically and culturally appropriate. Both can
          be drafted in a few days by one clinician and one cultural reviewer.
    \item \textbf{Adapt the configuration files.} Update subdomain topic lists,
          prompt styles, and variation settings to reflect local vocabulary and
          developmental priorities. This is editing, not engineering.
    \item \textbf{Run the SDE pipelines in small batches and review.} Generate
          10--20 questions and images, then have clinicians review them. The
          multi-judge system catches many problems automatically, but early human
          review is what reveals whether the cultural guidelines are specific
          enough.
    \item \textbf{Localize the caregiver assistant and UI.} Rewrite the chatbot
          system prompt and translate interface strings. The assistant's
          tool-calling logic and database queries remain untouched.
\end{enumerate}

The underlying principle is that the system is designed so that clinical and
cultural adaptation happens in content files and configuration, not in
application code. A team replicating \textit{Digital Harf} for a new language does not
need to modify the therapy interaction logic, the personalization algorithms,
the SDE pipeline architecture, or the database schema. What they need is a
clinician who understands local therapy practice and a cultural reviewer who
can articulate visual and behavioral norms clearly enough for an image
generation model to follow them.
\section{Conclusion}

We presented \textit{Digital Harf}, a multimodal AI platform designed to extend speech
and language therapy into the home for Arabic-speaking children with ASD.
Developed through iterative collaboration with SLPs across the Middle East,
the system integrates structured therapeutic modules, adaptive image selection, and an AI-driven caregiver assistant into a unified workflow.
Its Agentic Synthetic Data Engine enables scalable generation of culturally
grounded therapy materials, directly addressing the content gap that limits
existing digital tools in Arabic-speaking and other non-English settings.

Expert evaluation demonstrated strong cultural and linguistic alignment, high
clinical acceptability of generated content (90.1\% acceptance without manual
editing), and promising usability. The evaluation also surfaced important
tensions around evidence-based validation---tensions the team is actively
addressing through a structured clinical validation study now underway across
multiple therapy centers, hospitals, and medical institutions in the region.
\textit{Digital Harf} offers early evidence that pervasive, AI-driven therapeutic systems can be designed from the ground up for underrepresented populations — treating cultural grounding as core infrastructure rather than an afterthought, and positioning caregivers and clinicians as collaborators rather than bystanders in technology-mediated care.

\begin{acks}
The authors gratefully acknowledge the 13 licensed Speech-Language Pathologists
who participated in the expert evaluation studies across three clinical centers
in Saudi Arabia: \textit{King Fahad Armed Forces Hospital},
\textit{Mohammed Bin Salman Autism Program}, and
\textit{Madinah Autism Society}. Their clinical expertise and feedback were
essential to the validation of this work.

We also thank Sama Almuraykhi and Sara Alghamdi from the Ministry of Defense
Digital Transformation, Saudi Arabia, for their support throughout this
research. Finally, we express our gratitude to Dr.~Waleed Alhazzani and
Dr.~Hadeel Abdulmohsen AlKhamees for their strategic guidance and sustained
engagement with this project.
\end{acks}

\bibliographystyle{ACM-Reference-Format}
\bibliography{sample-base}

@String{Computing = "Computing" }

@article{khamees2025estimating,
  author  = {Khamees, H. and Gulati, A. and Vatsa, A. and Aldosari, M. and Hoque, E.},
  title   = {Estimating autism prevalence among children in the Kingdom of Saudi Arabia: A comprehensive multisource benchmarking approach},
  journal = {Saudi Medical Journal},
  year    = {2025},
  volume  = {46},
  number  = {11},
  pages   = {1276--1279},
  month   = {November},
  doi     = {10.15537/smj.2025.46.11.20250781},
  pmid    = {41224349},
  pmcid   = {PMC12625678}
}

@article{azad2026harf,
  title   = {Harf-Speech: A Clinically Aligned Framework for Arabic Phoneme-Level Speech Assessment},
  author  = {Azad, Asif and Shanto, MD and Hossain, Mohammad Sadat and Alwuqaysi, Bdour and Boughorbel, Sabri and Bokhari, Yahya and Aljouie, Abdulrhman and Sindi, Ayah Othman and Hoque, Ehsan},
  journal = {arXiv preprint arXiv:2604.06191},
  year    = {2026}
}

@article{zajc2018tablet,
  title   = {Tablet game-supported speech therapy embedded in children's popular practices},
  author  = {Zajc, Matej and Isteni{\v c} Star{\v c}i{\v c}, Andreja and Lebeni{\v c}nik, Maja and Ga{\v c}nik, Mateja},
  journal = {Behaviour \& Information Technology},
  volume  = {37},
  number  = {7},
  pages   = {693--702},
  year    = {2018},
  doi     = {10.1080/0144929X.2018.1474253}
}

@article{hair2021apraxia,
  title   = {A Longitudinal Evaluation of Tablet-Based Child Speech Therapy with {Apraxia World}},
  author  = {Hair, Adam and Ballard, Kirrie J. and Markoulli, Constantina and Mckechnie, Jacqueline and Monroe, Penelope and Ahmed, Beena and Gutierrez-Osuna, Ricardo},
  journal = {ACM Transactions on Accessible Computing},
  volume  = {14},
  number  = {1},
  pages   = {1--26},
  year    = {2021},
  doi     = {10.1145/3433607}
}

@article{deka2025aistreview,
  title   = {{AI}-based automated speech therapy tools for persons with speech sound disorder: a systematic literature review},
  author  = {Deka, Chinmoy and Shrivastava, Abhishek and Abraham, Ajish K. and Nautiyal, Saurabh and Chauhan, Praveen},
  journal = {Speech, Language and Hearing},
  volume  = {28},
  number  = {1},
  pages   = {2359274},
  year    = {2025},
  doi     = {10.1080/2050571X.2024.2359274}
}

@article{moulaei2025telespeech,
  title   = {Exploring tele-speech therapy: A scoping review of interventions, applications, benefits, and challenges},
  author  = {Moulaei, Khadijeh and Dinari, Fatemeh and Hosseini, Mobina and Almasi, Sohrab and Sabet, Babak and Anabestani, Romina and Afrash, Mohammad Reza},
  journal = {International Journal of Medical Informatics},
  volume  = {195},
  pages   = {105784},
  year    = {2025},
  doi     = {10.1016/j.ijmedinf.2025.105784}
}

@article{porayskapomsta2018blending,
  title   = {Blending Human and Artificial Intelligence to Support Autistic Children's Social Communication Skills},
  author  = {Porayska-Pomsta, Ka{\'s}ka and Alcorn, Alyssa M. and Avramides, Katerina and Beale, Sandra and Bernardini, Sara and Foster, Mary Ellen and Frauenberger, Christopher and Good, Judith and Guldberg, Karen and Keay-Bright, Wendy and Kossyvaki, Lila and Lemon, Oliver and Mademtzi, Marilena and Menzies, Rachel and Pain, Helen and Rajendran, Gnanathusharan and Waller, Annalu and Wass, Sam and Smith, Tim J.},
  journal = {ACM Transactions on Computer-Human Interaction},
  volume  = {25},
  number  = {6},
  pages   = {1--35},
  year    = {2018},
  doi     = {10.1145/3271484}
}

@inproceedings{choi2025aacess,
  title     = {{AACessTalk}: Fostering Communication between Minimally Verbal Autistic Children and Parents with Contextual Guidance and Card Recommendation},
  author    = {Choi, Dasom and Park, SoHyun and Lee, Kyungah and Hong, Hwajung and Kim, Young-Ho},
  booktitle = {Proceedings of the 2025 CHI Conference on Human Factors in Computing Systems},
  pages     = {1--25},
  publisher = {Association for Computing Machinery},
  address   = {New York, NY, USA},
  year      = {2025},
  doi       = {10.1145/3706598.3713792}
}

@inproceedings{suh2024opportunities,
  title     = {Opportunities and Challenges for {AI}-Based Support for Speech-Language Pathologists},
  author    = {Suh, Hyewon and Dangol, Aayushi and Meadan, Hedda and Miller, Carol A. and Kientz, Julie A.},
  booktitle = {Proceedings of the 3rd Annual Meeting of the Symposium on Human-Computer Interaction for Work},
  pages     = {1--14},
  publisher = {Association for Computing Machinery},
  address   = {New York, NY, USA},
  year      = {2024},
  doi       = {10.1145/3663384.3663387}
}

@inproceedings{dangol2025iwantslp,
  title     = {{``I Want to Think Like an SLP'': A Design Exploration of AI-Supported Home Practice in Speech Therapy}},
  author    = {Dangol, Aayushi and Lewis, Aaleyah and Suh, Hyewon and Hong, Xuesi and Meadan, Hedda and Fogarty, James and Kientz, Julie A.},
  booktitle = {Proceedings of the 2025 CHI Conference on Human Factors in Computing Systems},
  pages     = {1--22},
  publisher = {Association for Computing Machinery},
  address   = {New York, NY, USA},
  year      = {2025},
  doi       = {10.1145/3706598.3713986}
}

@inproceedings{lewis2025cld,
  title     = {Exploring {AI}-Based Support in Speech-Language Pathology for Culturally and Linguistically Diverse Children},
  author    = {Lewis, Aaleyah and Dangol, Aayushi and Suh, Hyewon and Olszewski, Abbie and Fogarty, James and Kientz, Julie A.},
  booktitle = {Proceedings of the 2025 CHI Conference on Human Factors in Computing Systems},
  pages     = {1--19},
  publisher = {Association for Computing Machinery},
  address   = {New York, NY, USA},
  year      = {2025},
  doi       = {10.1145/3706598.3714131}
}

@article{chueh2025genai,
  author    = {Chueh, C. H. and Chiang, T. H. and Pan, P. W. and Lin, K. L. and Lu, Y. S. and Tuan, S. H. and Lin, C. R. and Huang, I. C. and Cheng, H. S.},
  title     = {Implementation of a Generative AI-Powered Digital Interactive Platform for Clinical Language Therapy in Children with Language Delay: A Pilot Study},
  journal   = {Life},
  year      = {2025},
  volume    = {15},
  number    = {10},
  pages     = {1628},
  month     = {October},
  doi       = {10.3390/life15101628},
  pmid      = {41157300},
  pmcid     = {PMC12565666},
  publisher = {MDPI}
}

@article{attwell2022ostreview,
  title   = {A Systematic Review of Online Speech Therapy Systems for Intervention in Childhood Speech Communication Disorders},
  author  = {Attwell, Geertruida Aline and Bennin, Kwabena Ebo and Tekinerdogan, Bedir},
  journal = {Sensors},
  volume  = {22},
  number  = {24},
  pages   = {9713},
  year    = {2022},
  doi     = {10.3390/s22249713}
}

@article{almurashi2022asdreview,
  title   = {Augmented Reality, Serious Games and Picture Exchange Communication System for People with {ASD}: Systematic Literature Review and Future Directions},
  author  = {Almurashi, Haneen and Bouaziz, Rahma and Alharthi, Wallaa and Al-Sarem, Mohammed and Hadwan, Mohammed and Kammoun, Slim},
  journal = {Sensors},
  volume  = {22},
  number  = {3},
  pages   = {1250},
  year    = {2022},
  doi     = {10.3390/s22031250}
}

@article{lyu2023eggly,
  author     = {Lyu, Yue and An, Pengcheng and Xiao, Yage and Zhang, Zibo and Zhang, Huan and Katsuragawa, Keiko and Zhao, Jian},
  title      = {Eggly: Designing Mobile Augmented Reality Neurofeedback Training Games for Children with Autism Spectrum Disorder},
  year       = {2023},
  issue_date = {June 2023},
  publisher  = {Association for Computing Machinery},
  address    = {New York, NY, USA},
  volume     = {7},
  number     = {2},
  url        = {https://doi.org/10.1145/3596251},
  doi        = {10.1145/3596251},
  journal    = {Proc. ACM Interact. Mob. Wearable Ubiquitous Technol.},
  month      = jun,
  articleno  = {67},
  numpages   = {29}
}

@article{lyu2024emooly,
  title   = {{EMooly}: Supporting Autistic Children in Collaborative Social-Emotional Learning with Caregiver Participation through Interactive {AI}-Infused and {AR} Activities},
  author  = {Lyu, Yue and Liu, Di and An, Pengcheng and Tong, Xin and Zhang, Huan and Katsuragawa, Keiko and Zhao, Jian},
  journal = {Proceedings of the ACM on Interactive, Mobile, Wearable and Ubiquitous Technologies},
  volume  = {8},
  number  = {4},
  pages   = {1--36},
  year    = {2024},
  doi     = {10.1145/3699738}
}

@inproceedings{marcu2012wearablecameras,
  title     = {Parent-Driven Use of Wearable Cameras for Autism Support: A Field Study with Families},
  author    = {Marcu, Gabriela and Dey, Anind K. and Kiesler, Sara},
  booktitle = {Proceedings of the 2012 ACM Conference on Ubiquitous Computing},
  pages     = {401--410},
  publisher = {Association for Computing Machinery},
  address   = {New York, NY, USA},
  year      = {2012},
  doi       = {10.1145/2370216.2370277}
}

@article{koronthaly2025coevolution,
  title   = {The Co-Evolution of Language, Technology, and Culture in Autism Computing Research: Lessons Learned From {ACM} Digital Library},
  author  = {Koronth\'{a}ly, Daniel and Heffelman, Adrian and Magsi, Hamza and Zuber, Wilson and Sharmin, Moushumi and Ahmed, Shameem},
  journal = {Proceedings of the ACM on Interactive, Mobile, Wearable and Ubiquitous Technologies},
  volume  = {9},
  number  = {4},
  pages   = {1--31},
  year    = {2025},
  doi     = {10.1145/3770638}
}

@inproceedings{lai2025asdillm,
  title     = {{ASD}-i{LLM}:An Intervention Large Language Model for Autistic Children based on Real Clinical Dialogue Intervention Dataset},
  author    = {Lai, Shuzhong  and
               Li, Chenxi  and
               Lai, Junhong  and
               Zhong, Yucun  and
               Yan, Chenyu  and
               Li, Xiang  and
               Li, Haifeng  and
               Pan, Gang  and
               Yao, Lin  and
               Wang, Yueming},
  editor    = {Christodoulopoulos, Christos  and
               Chakraborty, Tanmoy  and
               Rose, Carolyn  and
               Peng, Violet},
  booktitle = {Findings of the Association for Computational Linguistics: EMNLP 2025},
  month     = nov,
  year      = {2025},
  address   = {Suzhou, China},
  publisher = {Association for Computational Linguistics},
  url       = {https://aclanthology.org/2025.findings-emnlp.427/},
  doi       = {10.18653/v1/2025.findings-emnlp.427},
  pages     = {8058--8079},
  isbn      = {979-8-89176-335-7}
}

@inproceedings{choi2024unlocklife,
  author    = {Choi, Dasom and Lee, Sunok and Kim, Sung-In and Lee, Kyungah and Yoo, Hee Jeong and Lee, Sangsu and Hong, Hwajung},
  title     = {Unlock Life with a Chat(GPT): Integrating Conversational AI with Large Language Models into Everyday Lives of Autistic Individuals},
  year      = {2024},
  isbn      = {9798400703300},
  publisher = {Association for Computing Machinery},
  address   = {New York, NY, USA},
  url       = {https://doi.org/10.1145/3613904.3641989},
  doi       = {10.1145/3613904.3641989},
  booktitle = {Proceedings of the 2024 CHI Conference on Human Factors in Computing Systems},
  articleno = {72},
  numpages  = {17},
  location  = {Honolulu, HI, USA},
  series    = {CHI '24}
}

@article{perry2024aireview,
  title   = {{AI} technology to support adaptive functioning in neurodevelopmental conditions in everyday environments: a systematic review},
  author  = {Perry, Nina and Sun, Calum and Munro, Mark and Boulton, K. A. and Guastella, Adam J.},
  journal = {npj Digital Medicine},
  volume  = {7},
  pages   = {362},
  year    = {2024},
  doi     = {10.1038/s41746-024-01355-7}
}

@article{deka2025hcai,
  author   = {Deka, Chinmoy and Shrivastava, Abhishek and Kumar, Rishav},
  title    = {Towards human-centered AI in speech therapy: perspectives from a low-resource setting},
  journal  = {Universal Access in the Information Society},
  year     = {2025},
  volume   = {25},
  number   = {1},
  pages    = {16},
  month    = {December},
  doi      = {10.1007/s10209-025-01288-2},
  url      = {https://doi.org/10.1007/s10209-025-01288-2},
  issn     = {1615-5297}
}

@inbook{baird2024asrdtx,
  author = {Lee, Seonwoo and Mun, Jihyun and Kim, Sunhee and Park, HyunJu and Yang, Suvin and Kim, HyunDon and Noh, SeungJae and Kim, WonBin and Chung, Minhwa},
  year   = {2024},
  month  = {07},
  pages  = {328-335},
  title  = {Automatic Speech Recognition and Assessment Systems Incorporated into Digital Therapeutics for Children with Autism Spectrum Disorder},
  isbn   = {978-3-031-62848-1},
  doi    = {10.1007/978-3-031-62849-8_40}
}

@inproceedings{attia2024kidwhisper,
  title     = {Kid-Whisper: Towards Bridging the Performance Gap in Automatic Speech Recognition for Children VS. Adults},
  author    = {Attia, Ahmed Adel and Liu, Jing and Ai, Wei and Demszky, Dorottya and Espy-Wilson, Carol},
  booktitle = {Proceedings of the AAAI/ACM Conference on AI, Ethics, and Society},
  volume    = {7},
  pages     = {74--80},
  year      = {2024},
  doi       = {10.1609/aies.v7i1.31618}
}

@article{kim2025koreassr,
  title   = {Usefulness of Automatic Speech Recognition Assessment of Children With Speech Sound Disorders: Validation Study},
  author  = {Kim, Do Hyung and Jeong, Joo Won and Kang, Dayoung and Ahn, Taekyung and Hong, Yeonjung and Im, Younggon and Kim, Jaewon and Kim, Min Jung and Jang, Dae-Hyun},
  journal = {Journal of Medical Internet Research},
  volume  = {27},
  pages   = {e60520},
  year    = {2025},
  doi     = {10.2196/60520}
}

@article{atturu2025cognitivebotics,
  title   = {Effectiveness of Artificial Intelligence--Based Platform in Administering Therapies for Children With Autism Spectrum Disorder: 12-Month Observational Study},
  author  = {Atturu, Harini and Naraganti, Somasekhar and Rao, Bugatha Rajvir},
  journal = {JMIR Neurotechnology},
  volume  = {4},
  pages   = {e70589},
  year    = {2025},
  doi     = {10.2196/70589}
}

@misc{PAELS_TPT,
  author       = {{Speech and Language at Home}},
  title        = {Photograph Assessment of Early Language Skills: Early Childhood Autism Speech},
  year         = {n.d.},
  howpublished = {\url{https://www.teacherspayteachers.com/Product/Photograph-Assessment-of-Early-Language-Skills-Early-Childhood-Autism-Speech-921161}},
  note         = {Teachers Pay Teachers resource}
}

@misc{kscdr2024stats,
  author       = {{King Salman Center for Disability Research}},
  title        = {Disability Statistics in the Kingdom of {Saudi Arabia}},
  year         = {2024},
  howpublished = {\url{https://www.kscdr.org.sa/en/stats}},
  note         = {Accessed: September 11, 2024}
}

@article{tinelli2023economic,
  title   = {Economic evaluation of anti-epileptic medicines for autistic children with epilepsy},
  author  = {Tinelli, Michela M. and Roddy, Aine and Knapp, Martin and others},
  journal = {Journal of Autism and Developmental Disorders},
  year    = {2023},
  doi     = {10.1007/s10803-023-05941-8}
}

@article{ghali2023audiology,
  author  = {Ghali, S. and Abdalla, F. and Aseeri, A.},
  title   = {Opportunities and Challenges for Audiology and Speech-Language Pathology Services in Arabic-Speaking Middle East Countries},
  journal = {Perspectives of the ASHA Special Interest Groups},
  year    = {2023},
  volume  = {8},
  number  = {1},
  pages   = {217--225},
  doi     = {10.1044/2022_persp-22-00103},
  pmid    = {39639998},
  pmcid   = {PMC11619755},
  url     = {https://doi.org/10.1044/2022_persp-22-00103}
}

@misc{kidzcare2025app,
  author = {{KidzCare Pediatrics}},
  title  = {KidzCare App},
  year   = {2025},
  url    = {https://kidzcare.app/en/},
  note   = {Accessed: 2026-05-01}
}

@misc{microsoft_pronunciation_assessment,
  author       = {{Microsoft}},
  title        = {Use pronunciation assessment},
  url          = {https://learn.microsoft.com/en-us/azure/ai-services/speech-service/how-to-pronunciation-assessment?pivots=programming-language-python},
  note         = {Microsoft Learn documentation. Accessed: May 2, 2026}
}

\appendix

\section{Supporting Detail for the Studies}\label{sec:appendix-supporting-detail}

This appendix provides the instrument structure used in both expert studies and
the detailed tables that support the focused findings in the main text.

\subsection{Study 1 Content-Validation Instrument}\label{sec:appendix-study1-asde-contents}

The Study 1 questionnaire embedded the sampled therapy items directly in the
questionnaire. Participants made a binary acceptability judgement for each of
the 27~items (12~Picture Description and 15~Language Therapy).
Table~\ref{tab:appendix-study1-items} reports item-level acceptability in the
survey.

\begin{figure}[htbp]
    \centering
    \includegraphics[width=\linewidth,height=0.88\textheight,keepaspectratio]%
    {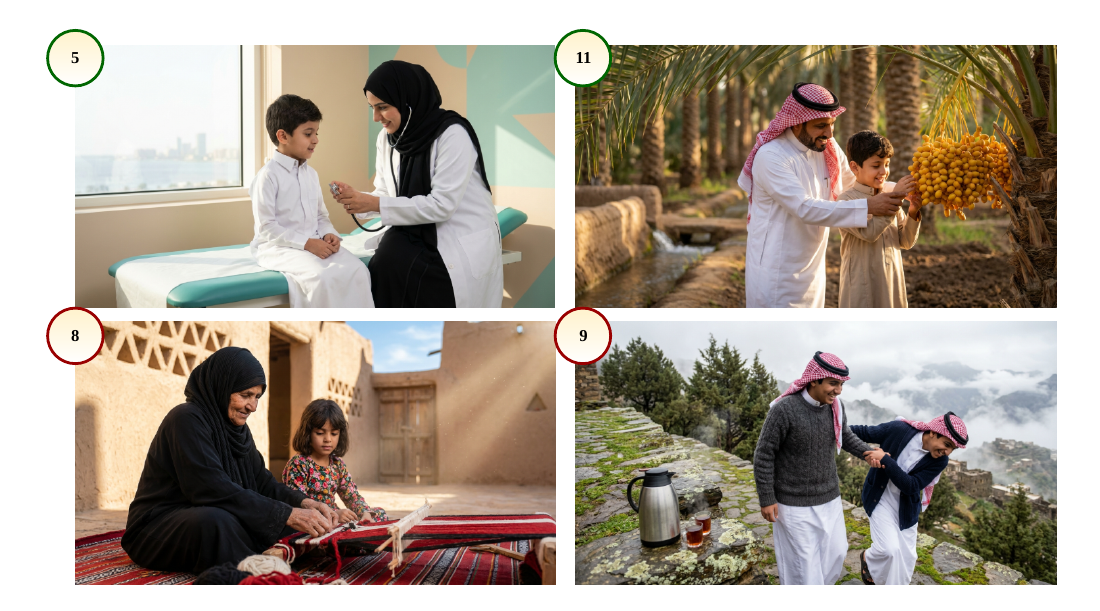}
    \caption{%
        Examples of Picture Description content generated by ASDE and shown to
        SLPs in Study~1.
        \textbf{Top two:} highest-rated items (PD-5 and PD-11).
        \textbf{Bottom two:} lowest-rated items (PD-8 and PD-9);
        see Table~\ref{tab:appendix-study1-items}.}
    \label{fig:appendix-study1-pd-asde}
\end{figure}

\begin{figure}[htbp]
    \centering
    \includegraphics[width=\linewidth,height=0.88\textheight,keepaspectratio]%
    {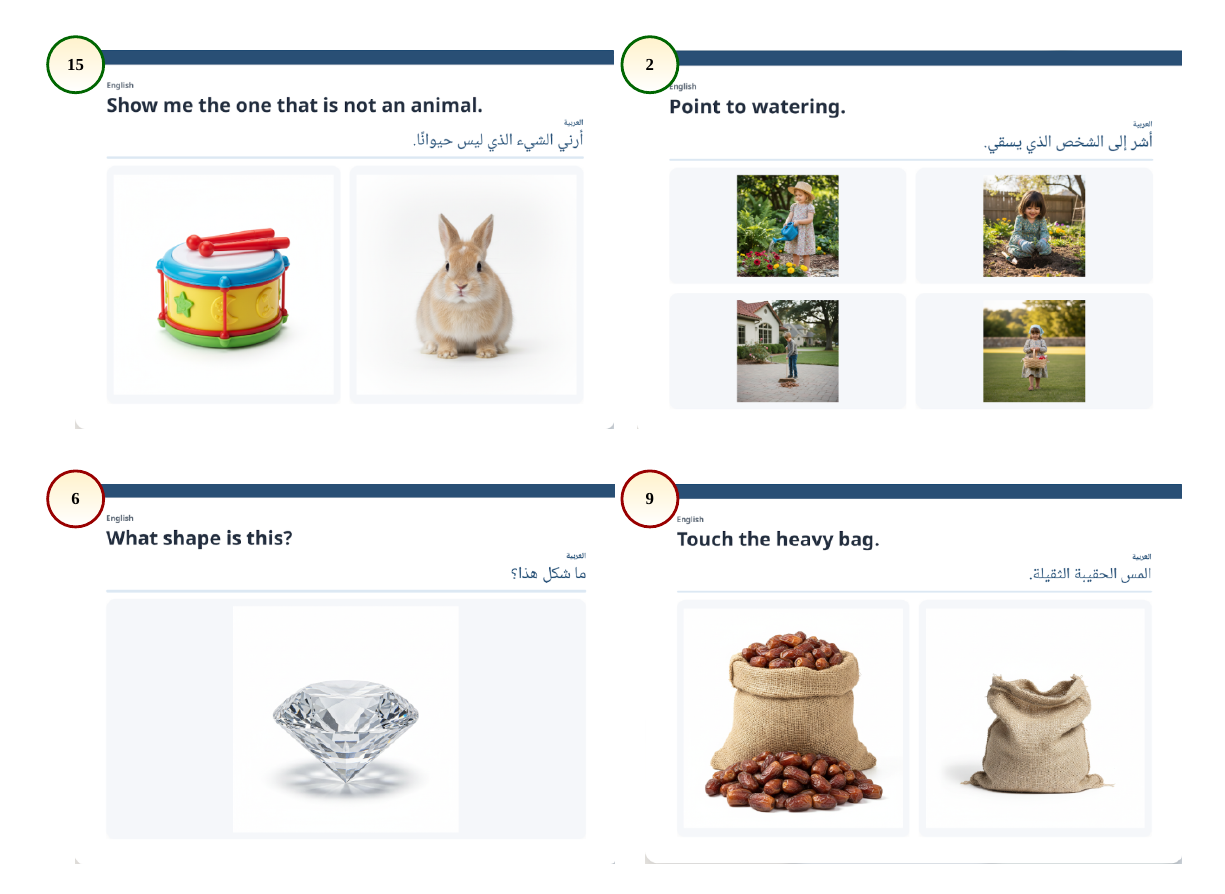}
    \caption{%
        Examples of Language Therapy content generated by ASDE and shown to SLPs
        in Study~1.
        \textbf{Top two:} highest-rated items---LT-15 (Receptive, Negative domain)
        and LT-2 (Receptive, Common Actions domain).
        \textbf{Bottom two:} lowest-rated items---LT-6 (Expressive, Shape domain)
        and LT-9 (Receptive, Opposites domain);
        see Table~\ref{tab:appendix-study1-items}.}
    \label{fig:appendix-study1-lt-asde}
\end{figure}

Figures~\ref{fig:appendix-study1-pd-asde} and~\ref{fig:appendix-study1-lt-asde}
present representative ASDE-generated content shown to the SLPs during the
Study~1 survey. In both figures the top two examples are drawn from the
highest-rated items and the bottom two from the lowest-rated items listed in
Table~\ref{tab:appendix-study1-items}.

\begin{table}[htbp]
    \sffamily\small
    \def\arraystretch{1.02}\setlength{\tabcolsep}{0.26em}
    \centering
    \arrayrulecolor{appendixrule}
    \caption{%
        Item-level acceptability rates from Study~1.
        ``Accepted'' = number of the nine SLPs who marked the item suitable
        for use without editing.
        Items are grouped by module: PD (Picture Description) and
        LT (Language Therapy).
        Overall acceptance across 243~judgements ($9{\times}27$) was 90.1\%;
        LT (93.3\%) outperformed PD (86.1\%).
        No item fell below two-thirds acceptance.}
    \label{tab:appendix-study1-items}
    \begin{tabular}{|L{0.18\textwidth}!{\color{appendixrule}\vrule}C{0.11\textwidth}!{\color{appendixrule}\vrule}%
        C{0.13\textwidth}!{\color{appendixrule}\vrule}C{0.12\textwidth}|}
        \hline
        \rowcolor{appendixheader}
        \textbf{Module}      & \textbf{Item}     & \textbf{Accepted} & \textbf{Rate (\%)} \\
        \hline
        \rowcolor{appendixsubheader}
        \multicolumn{4}{|l|}{\textbf{Picture Description (PD)}}                           \\
        \hline
        PD                   & PD-1              & 7/9               & 77.8               \\
        PD                   & PD-2              & 7/9               & 77.8               \\
        PD                   & PD-3              & 8/9               & 88.9               \\
        PD                   & PD-4              & 8/9               & 88.9               \\
        PD                   & PD-5              & 9/9               & 100.0              \\
        PD                   & PD-6              & 7/9               & 77.8               \\
        PD                   & PD-7              & 9/9               & 100.0              \\
        PD                   & PD-8              & 6/9               & 66.7               \\
        PD                   & PD-9              & 6/9               & 66.7               \\
        PD                   & PD-10             & 8/9               & 88.9               \\
        PD                   & PD-11             & 9/9               & 100.0              \\
        PD                   & PD-12             & 9/9               & 100.0              \\
        \rowcolor{appendixpanel}
        \textbf{PD subtotal} & ---               & \textbf{93/108}   & \textbf{86.1}      \\
        \hline
        \rowcolor{appendixsubheader}
        \multicolumn{4}{|l|}{\textbf{Language Therapy (LT)}}                              \\
        \hline
        LT                   & LT-1              & 8/9               & 88.9               \\
        LT                   & LT-2              & 9/9               & 100.0              \\
        LT                   & LT-3              & 9/9               & 100.0              \\
        LT                   & LT-4              & 9/9               & 100.0              \\
        LT                   & LT-5              & 9/9               & 100.0              \\
        LT                   & LT-6              & 7/9               & 77.8               \\
        LT                   & LT-7              & 8/9               & 88.9               \\
        LT                   & LT-8              & 9/9               & 100.0              \\
        LT                   & LT-9              & 7/9               & 77.8               \\
        LT                   & LT-10             & 9/9               & 100.0              \\
        LT                   & LT-11             & 9/9               & 100.0              \\
        LT                   & LT-12             & 8/9               & 88.9               \\
        LT                   & LT-13             & 8/9               & 88.9               \\
        LT                   & LT-14             & 8/9               & 88.9               \\
        LT                   & LT-15             & 9/9               & 100.0              \\
        \rowcolor{appendixpanel}
        \textbf{LT subtotal} & ---               & \textbf{126/135}  & \textbf{93.3}      \\
        \hline
        \rowcolor{appendixheader}
        \textbf{Combined}    & \textbf{27 items} & \textbf{219/243}  & \textbf{90.1}      \\
        \hline
    \end{tabular}
    \arrayrulecolor{black}
\end{table}

\section{Technical Details}\label{sec:appendix-technical}

This section groups backend mechanisms that are summarized briefly in the main
text but are central to how \textit{Digital Harf} adapts Picture Description
content and grounds caregiver support.

\subsection{Score-Aware Matching (SAM) Algorithm}\label{sec:appendix-sam}

Algorithm~\ref{alg:sam} gives the pseudocode for the Score-Aware Matching (SAM)
algorithm that powers the Adaptive Image Selector in the Picture Description
module. The algorithm operates in two phases: a \emph{selection phase} executed
on session start (or new-image request) that chooses the next image to present,
and a \emph{logging phase} executed after the user responds that persists
session data for future selections. The frontend resolves a three-level
fallback chain---backend SAM selection, then database-random, then a hardcoded
emergency image---to ensure the module remains operational under transient
backend failures.

\begin{algorithm}[htbp]
    \caption{Score-Aware Matching (SAM) for Adaptive Image Selection}
    \label{alg:sam}
    \small
    \begin{algorithmic}[1]
        \Require Authenticated user $u$;\;difficulty level $d$ (from UI);\;
        recency window $W{=}5$
        \Ensure  Selected image $I^{*}$ with pool metadata

        \Statex \textbf{--- Frontend: on page load / new-image request ---}
        \State $d \gets$ difficulty level chosen in the UI
        \State $I^{*} \gets$ \Call{BackendSelect}{$u$, $d$}
        \If{$I^{*}$ is \textbf{null}}
        \State $I^{*} \gets$ \Call{DatabaseRandom}{$d$} \Comment{fallback 1}
        \EndIf
        \If{$I^{*}$ is \textbf{null}}
        \State $I^{*} \gets$ \Call{HardcodedEmergencyImage}{} \Comment{fallback 2}
        \EndIf

        \Statex
        \Statex \textbf{--- Backend: \Call{BackendSelect}{$u$, $d$} ---}
        \State $H \gets$ all past Picture Description sessions for user $u$
        \State $\overline{s}[i] \gets \mathrm{mean}\bigl(\{s:(i,s)\in H\}\bigr)$
        for each image $i$ seen by $u$
        \State $\mathrm{seen} \gets \{i : i \text{ appears in } H\}$

        \Statex \textit{// Partition images into performance pools}
        \State $P_{\mathrm{init}} \gets \{i \in \mathcal{I} :
            i \notin \mathrm{seen}\}$ \Comment{never scored}
        \State $P_{\mathrm{hard}} \gets$ lowest-scoring subset of $\mathrm{seen}$
        \State $P_{\mathrm{med}}  \gets$ middle-scoring subset of $\mathrm{seen}$
        \State $P_{\mathrm{easy}} \gets$ highest-scoring subset of $\mathrm{seen}$

        \Statex \textit{// Recency filter: exclude images seen in last $W$ sessions}
        \State $\mathcal{R} \gets \{i : i \text{ appeared in last } W
            \text{ sessions of } u\}$
        \For{each pool $P \in \{P_{\mathrm{init}}, P_{\mathrm{hard}},
                P_{\mathrm{med}}, P_{\mathrm{easy}}\}$}
        \State $P \gets P \setminus \mathcal{R}$
        \EndFor

        \If{all pools are empty}
        \State Restore pools to pre-filter state and retry
        \Comment{relax recency}
        \EndIf
        \If{all pools still empty}
        \State \Return \textbf{null} \Comment{triggers frontend fallback}
        \EndIf

        \Statex \textit{// Weighted probabilistic selection}
        \State $(w_{\mathrm{init}}, w_{\mathrm{hard}}, w_{\mathrm{med}},
            w_{\mathrm{easy}}) \gets$ predefined weights;\;
        $w_{\mathrm{init}}$ is highest
        \If{$P_{\mathrm{init}} \neq \emptyset$}
        \State $I^{*} \gets \mathrm{uniformSample}(P_{\mathrm{init}})$
        \Else
        \State $I^{*} \gets \mathrm{weightedSample}(
            P_{\mathrm{hard}},P_{\mathrm{med}},P_{\mathrm{easy}}\,;\;
            w_{\mathrm{hard}},w_{\mathrm{med}},w_{\mathrm{easy}})$
        \EndIf
        \State \Return $I^{*}$ with metadata $\{\mathrm{pool},\,\mathrm{reason}\}$

        \Statex
        \Statex \textbf{--- After user responds ---}
        \State Log: $\{u,\;I^{*},\;\mathrm{score},\;
            \mathrm{transcription},\;\mathrm{duration}\}$
        \State Log per-dimension feedback categories
        \State Update $H$ so next \Call{BackendSelect}{$u$,$d$}
        reflects this session
    \end{algorithmic}
\end{algorithm}

\subsection{AI Caregiver Assistant}\label{sec:appendix-ai-assistant}

This subsection provides the full technical architecture of the AI Caregiver
Assistant, summarized in the main text (Section~\ref{sec:ai-assistant}).

\begin{figure*}[t]
    \centering
    \includegraphics[width=0.75\textwidth]{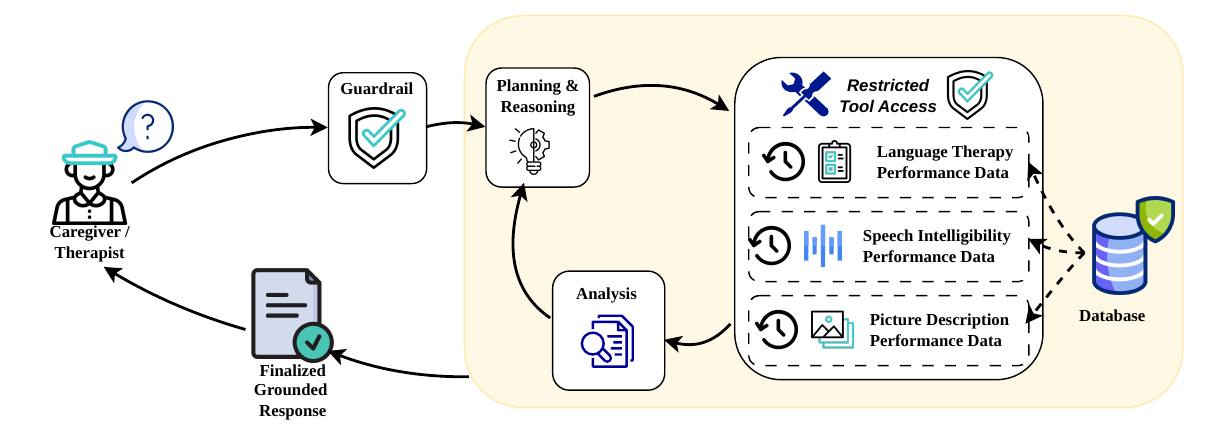}
    \caption{Technical workflow of the AI Caregiver Assistant. A caregiver query passes through a safety guardrail, then enters an Agentic Reasoning Core that retrieves performance data from the Language Therapy, Speech Intelligibility, and Picture Description modules. This data is synthesized into a personalized, evidence-grounded response. The architecture is modular, allowing different language models as the reasoning backend.}
    \label{fig:ai-technical}
    \Description{A flowchart showing the AI Caregiver Assistant pipeline. A query passes through a safety guardrail into an Agentic Reasoning Core, which connects to three module-specific data streams. The core synthesizes a personalized response grounded in the child's performance data.}
\end{figure*}

At the technical level (Figure~\ref{fig:ai-technical}), each query passes
through a safety guardrail before entering an \textit{Agentic Reasoning Core}.
The system plans how to answer the query and retrieves relevant data through
restricted access to module-specific performance streams (e.g., language
therapy, speech intelligibility, and picture description). This data is
analyzed and synthesized into a grounded response, ensuring that outputs are
directly supported by evidence rather than heuristic generation. The
architecture is modular and extensible, allowing integration of different
language models and retrieval systems while maintaining a consistent reasoning
pipeline. By combining conversational interaction with data-driven
personalization, the assistant acts as a bridge between clinical sessions and
continuous home-based support.

\section{User Interface and Interaction Design}\label{sec:appendix-ui}

This section presents the user-facing interface and interaction flow for each
of \textit{Digital Harf}'s therapeutic modules. Each module follows a
consistent design principle: simple navigation, multimodal prompts (text and
audio), and immediate, interpretable feedback at every step.

\subsection{Language Therapy Interface}\label{sec:appendix-lt-ui}

\begin{figure*}[htbp]
    \centering
    \includegraphics[width=\textwidth,height=0.85\textheight,keepaspectratio]{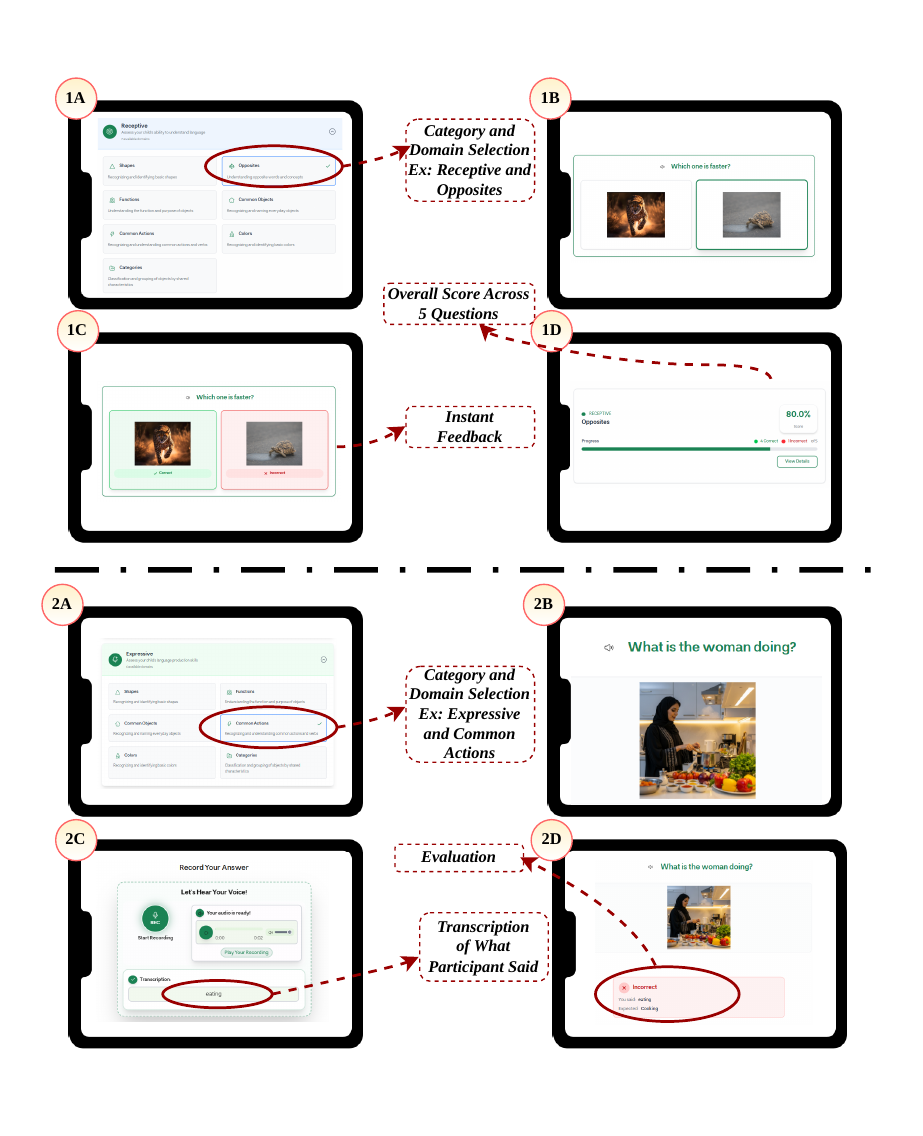}
    \caption{Language Therapy user interface and interaction flow, showing both Receptive and Expressive modes. \textbf{Receptive (top row):} (1A)~Mode and domain selection. (1B)~A question with text prompt, audio button, and image-based answer options. (1C)~Immediate correct/incorrect feedback. (1D)~Session score summary. \textbf{Expressive (bottom row):} (2A)~Mode and domain selection. (2B)~Question with audio prompt and supporting images. (2C)~The user records a spoken response, which is transcribed. (2D)~Semantic evaluation with immediate correctness feedback on the spoken response. Both modes use large tap targets and clear visual cues for children with limited reading ability.}
    \label{fig:appendix-lt-ui}
    \Description{A composite screenshot showing the Language Therapy interface across two rows. The top row shows Receptive mode in four panels: mode selection, a question with image options, correct-answer feedback, and a session score. The bottom row shows Expressive mode: mode selection, a question with audio prompt, a recording and transcription interface, and evaluation feedback on the spoken response.}
\end{figure*}

The Language Therapy module supports two interaction modes---Receptive and
Expressive---each following a structured session flow. In Receptive mode, the
child selects the correct image from multiple options in response to a text and
audio prompt. In Expressive mode, the child records a spoken answer, which the
system transcribes and evaluates. Both modes present five questions per domain,
provide immediate per-question feedback, and conclude with a session-level
score summary. Figure~\ref{fig:appendix-lt-ui} shows the complete interaction
sequence for both modes.

\subsection{Speech Intelligibility Interface}\label{sec:appendix-si-ui}

\begin{figure*}[htbp]
    \centering
    \includegraphics[width=\textwidth,height=0.95\textheight,keepaspectratio]{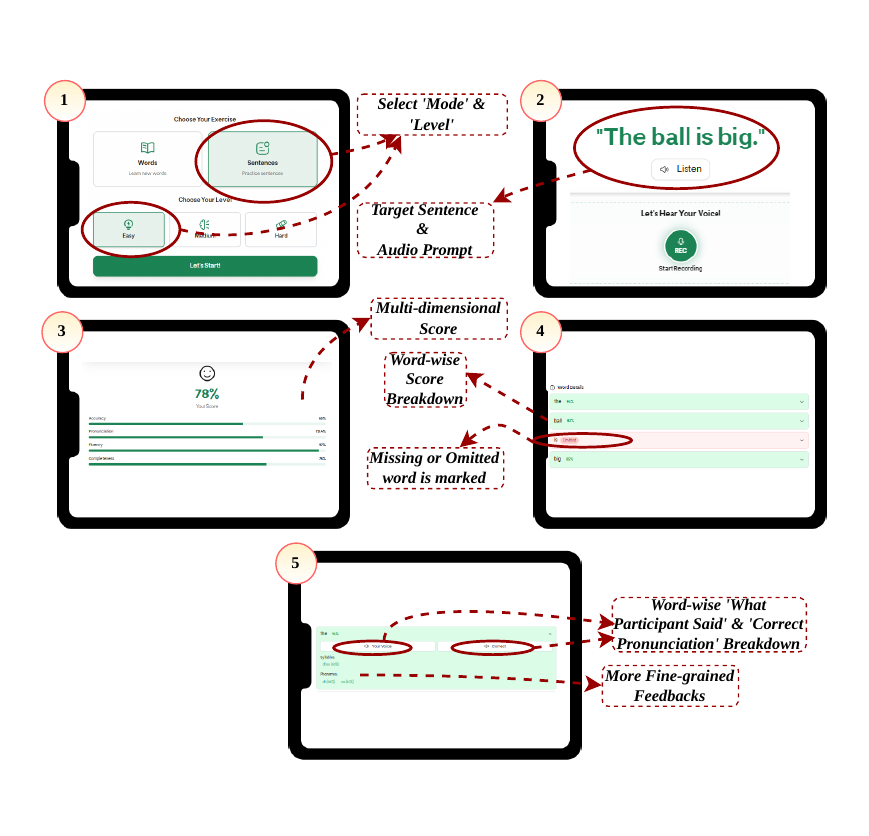}
    \caption{Speech Intelligibility user interface and interaction flow. (1)~The participant selects a mode and level. (2)~Based on the participant's history and past performance, five words or sentences are dynamically generated and presented one by one. For each item, the participant reads or listens to the prompt, attempts to utter it, and records the response. (3)~An instant evaluation score is provided across four dimensions. (4)~For sentence mode, word-level scores are also shown for each utterance. (5)~Additional fine-grained details can compare the spoken output with the target text and surface more specific pronunciation feedback when supported by the backend engine.}
    \label{fig:appendix-si-ui}
    \Description{A composite screenshot showing the Speech Intelligibility interface across the main stages of interaction, including activity selection, prompt presentation, speech recording, immediate feedback, and a results summary.}
\end{figure*}

The Speech Intelligibility module is designed around short, guided speaking
activities and instant evaluation to help children practice word- and
sentence-level production while receiving immediate feedback. In a typical
session, the child first selects a mode and level. Then, based on the child's
history and past mistakes, five words or sentences are presented one by one.
For each item, the child listens to or reads the prompt, records a spoken
response, and then receives intelligibility-related feedback together with a
session-level summary of the results. Figure~\ref{fig:appendix-si-ui} presents
an example of the user interface and interaction flow for this module.

\subsection{Picture Description Interface}\label{sec:appendix-pd-ui}

\begin{figure*}[htbp]
    \centering
    \includegraphics[width=\textwidth,height=0.95\textheight,keepaspectratio]{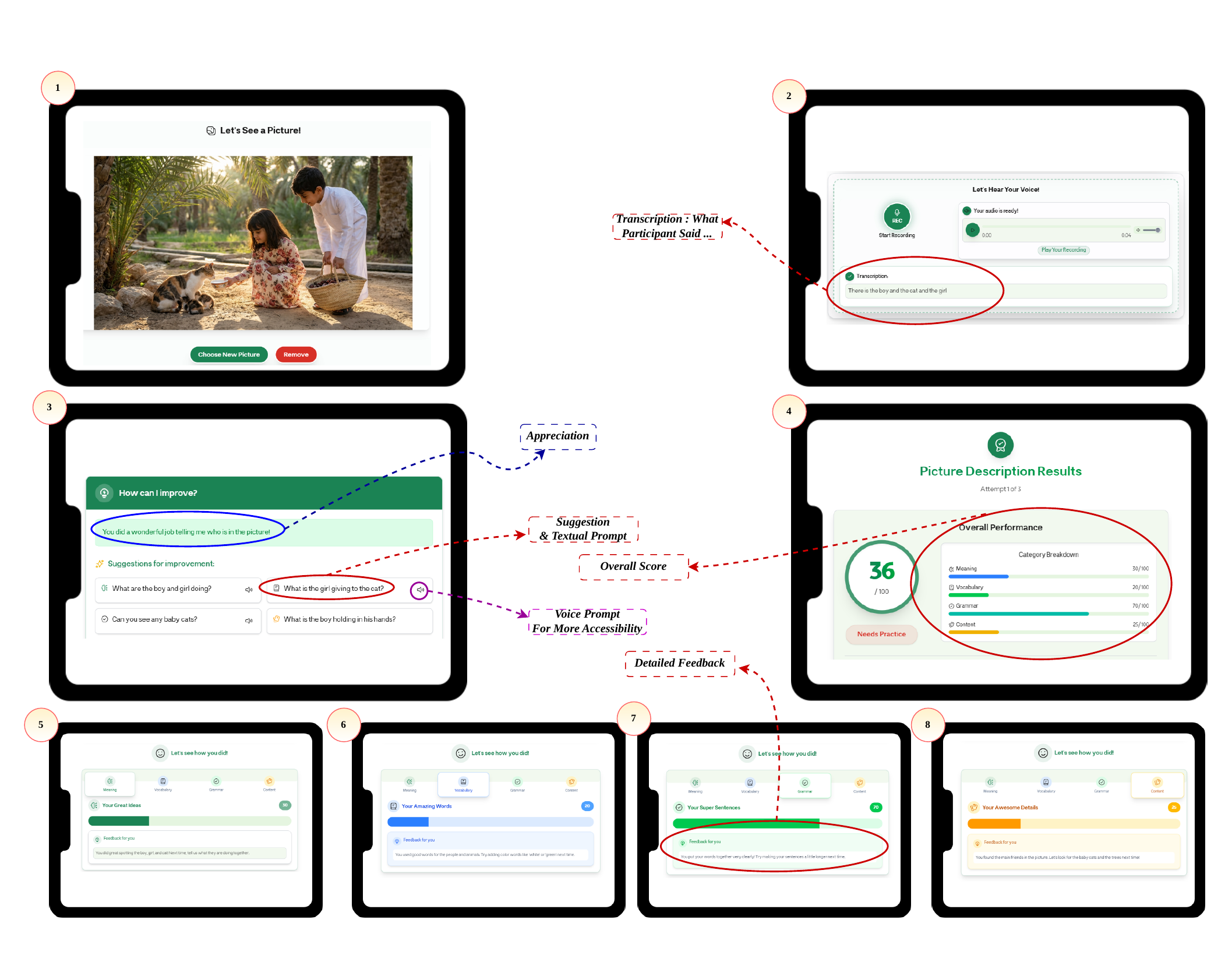}
    \caption{Picture Description user interface and interaction flow. (1)~A culturally relevant image is shown; the user may request a different one. (2)~The user records a spoken description and sees a transcription of the response. (3)~The system provides encouragement plus targeted follow-up prompts in text and audio to guide the next attempt. (4)~Overall performance is summarized with a total score and category breakdown. (5)--(8)~Detailed feedback cards present dimension-specific results for Meaning, Vocabulary, Grammar, and Content. This progressive disclosure lets caregivers decide how much detail to share with the child.}
    \label{fig:appendix-pd-ui}
    \Description{A composite screenshot showing the Picture Description interface in eight panels: an image display with a change button, a recording and transcription interface, a suggestion screen with encouraging and targeted prompts, an overall score summary with category breakdown, and four detailed feedback cards covering Meaning, Vocabulary, Grammar, and Content.}
\end{figure*}

The Picture Description module uses a multi-attempt interaction loop designed
to encourage progressively richer verbal descriptions. The child views a
culturally relevant image, records a spoken description, and receives targeted
suggestions for improvement. Up to three attempts are supported, with each
round of feedback building on the previous response. After the final attempt,
the system offers a detailed evaluation covering content completeness,
vocabulary quality, grammatical correctness, and semantic accuracy.
Figure~\ref{fig:appendix-pd-ui} presents the full interaction sequence.

\subsection{AI Caregiver Assistant Interaction}\label{sec:appendix-ai-ui}

Caregivers access the assistant through a bilingual chat interface and can ask
questions in Arabic or English. Unlike generic chatbots, the responses are not
purely general or template-based; instead, they are personalized and informed
by the child's performance across different modules. Queries about progress,
strengths, or areas of improvement are answered using the participant's
historical data, resulting in targeted, actionable recommendations rather than
generic advice. The assistant also supports general informational queries while
clearly indicating when professional consultation is required. The backend
reasoning architecture that supports this interaction is detailed in Appendix
Section~\ref{sec:appendix-ai-assistant} and Figure~\ref{fig:ai-technical}.

\end{document}